\documentclass[12pt]{iopart}
\usepackage{iopams}
\usepackage{graphicx}
\expandafter\let\csname equation*\endcsname\relax
\expandafter\let\csname endequation*\endcsname\relax
\usepackage{amsmath}
\usepackage{amssymb}
\usepackage{cite}
 
\usepackage{xcolor}

\begin{document}

\title[The distribution of eccentricities in random regular graphs]
{The distribution of eccentricities in random regular graphs  
}
\author{Dor Lev-Ari, Ofer Biham and Eytan Katzav}
\address{Racah Institute of Physics, The Hebrew University, Jerusalem 9190401, Israel}
\eads{\mailto{dor.lev-ari@mail.huji.ac.il},
\mailto{ofer.biham@mail.huji.ac.il}, 
\mailto{eytan.katzav@mail.huji.ac.il}}

\begin{abstract}

We derive a closed-form analytical expression for the distribution 
of eccentricities (DoE)  
in random regular graphs (RRGs) that consist of $N$ nodes of degree $c$.
The DoE is given by the tail distribution
$P(E > \ell) \simeq 1 - \exp \left[ - \exp  \left( - \frac{ e^{b \ell} - \mu }{\beta} \right) \right]$,
where the distance $\ell$ takes integer values,
$b = \ln (c-1)$ is the shape parameter,
$\beta = \frac{c-2}{c} N$ 
is the scale parameter and
$\mu =   \frac{c-2}{c} N  \ln N$
is the location parameter.
By providing the full distribution rather than a single characteristic length scale,
we present a detailed view of the large-scale structure.
In spite of the fact that the degrees of all the nodes are the same,
their eccentricities exhibit non-trivial variations.
We derive a closed-form expression for 
the mean eccentricity, 
which is given by
$\langle E \rangle \simeq   
\frac{\ln N}{\ln (c-1)} 
+
\frac{\ln \ln N}{\ln (c-1)} 
- \frac{  \ln c  -  \ln (c-2) }{ \ln (c-1) }
+ \frac{1}{2}$.
We calculate the
mode of the DoE,
which exhibits a staircase profile as a function of the network size.
Interestingly, the mode is given by
$E_{\rm mode} ={\rm Round} \left(  \langle E \rangle  \right)$,
where ${\rm Round}( x )$ is the nearest integer to $x$.
We also calculate the variance ${\rm Var}(E)$ and 
show that it exhibits oscillations as a function of the
network size $N$.
The results presented in this paper may serve as benchmarks for algorithmic approaches
to eccentricity calculations in large sparse networks.
The eccentricities are important in practical applications 
such as broadcasting and global dissemination,
where the network performance is determined by the longest delay times.

\end{abstract}


\maketitle

\section{Introduction}

Random networks (or graphs) are composed of $N$
nodes whose connections are established according to a stochastic rule.
They provide a theoretical framework for studying the 
structure and dynamics of a large variety of complex systems
\cite{Havlin2010,Newman2010,Dorogovtsev2022}.
A widely used approach for constructing random networks
with a prescribed degree distribution is the configuration model
\cite{Bollobas1980,Molloy1995,Molloy1998,Newman2001}.
In this construction, each node is assigned a degree drawn independently 
from a given degree distribution $P(k)$, 
generating a degree sequence $k_1,k_2,\dots,k_N$.
In the construction, each node $i$ is attached to $k_i$ half-edges or stubs,
which are randomly paired to stubs of other nodes to form the edges of the network.
The range of admissible degrees may be restricted to
$k_{\rm min} \le k \le k_{\rm max}$,
where $k_{\rm min}$ is the minimal allowed degree and $k_{\rm max}$ is the maximal allowed degree.
The mean degree $\langle K \rangle$ is denoted by $c$.
The configuration model generates maximum entropy ensembles 
in which the degree distribution $P(k)$ is fixed
\cite{Newman2001,Fronczak2004,Molloy1995,Molloy1998}.
A closely related canonical random-graph ensemble
is the Erd{\H o}s-R\'enyi (ER) network 
\cite{Erdos1959,Erdos1960,Erdos1961}, 
whose degree distribution converges to a Poisson distribution.
In fact, ER networks are special in the sense that it suffices to choose
the mean degree $c$ and the Poisson degree distribution emerges spontaneously
from the construction.

The random regular graph (RRG) is a special case of a configuration model network,
in which all the nodes are of the same degree $c$
\cite{Wormald1999}.
In other words, in
the case of an RRG 
the degree distribution is a degenerate
distribution of the form 
$P(k)=\delta_{k,c}$.
In this paper we focus on the case of $c \ge 3$.
To construct an RRG consisting of $N$ nodes of degree $c \ge 3$
(where $Nc$ is an even number),
we create a multiset of $Nc$ stubs which includes $c$ stubs for each node $i=1,2,\dots,N$.
Pairs of stubs are then selected randomly and connected to each other
to form edges between the corresponding nodes.
To illustrate the process we represent the stubs by $Nc$ balls, where the $c$ balls
associated with node $i$ are marked by $i$.
We then choose a random arrangement of the $Nc$ balls
in an array of $Nc/2$ cells, such that each cell includes exactly two balls.
In practice, a random arrangement of balls into cells can be obtained by generating 
a random permutation of the $Nc$ balls and grouping them sequentially into $Nc/2$
pairs, making the construction straightforward to implement.
A cell containing balls 
$i$ and $j$ represents an edge between nodes $i$ and $j$.
The representation in terms of balls and cells is particularly convenient for implementation
on the computer, since a single random permutation of the $Nc$ balls produces a 
uniformly random pairing of stubs, from which the network can be constructed directly.

The network obtained from the procedure described above is a multigraph,
which may include self-loops
(edges connecting a node to itself)
or multiple edges (two or more edges connecting the same pair of nodes).
To eliminate the self-loops and multiple edges, 
we apply an  
edge-switching process, 
which yields a simple graph while preserving the degree sequence.
In this process, 
as long as the network has not yet become a simple graph,
at each time step we select randomly one of the self-loops $(i,i)$ or one of the multiple edges
$(i,j)$. 
In case that a self-loop $(i,i)$ was selected, we choose
a random edge $(i',j')$ and swap the two edges into $(i,i')$ and $(i,j')$.
Similarly, in case that a multiple edge $(i,j)$ was selected, we
select a random edge $(i',j')$ and swap the two edges into $(i,i')$ and $(j,j')$.
In both cases, we complete the move only 
after we make sure that the swapping does not create a new self-loop or a new multiple edge. 
This random edge-switching process continues until no self-loops or multiple edges remain. 
The procedure described above provides the RRG ensemble used in the simulations.

A classical result of Wormald shows that 
in the limit of network size $N \rightarrow \infty$,
the probability $P_{\rm S}(N,c)$ that an RRG with a given degree $c \ge 3$
will consist of a single (S) connected component converges to $1$
\cite{Wormald1981}.
Using the terminology of percolation theory,
one can say that in the large network limit
the giant component encompasses the whole network.
In Appendix A we present an asymptotic expression
for the probability $P_{\rm M}(N,c) = 1 - P_{\rm S}(N,c)$ that an RRG of size
$N$ and degree $c$ will consist of multiple (M) components.
It is shown that this probability scales like
$P_{\rm M}(N,c) \sim  N^{- \frac{(c+1)(c-2)}{2}}$
and thus becomes negligible for sufficiently large networks.
Moreover, its decay rate is accelerated as $c$ is increased.
This implies that, in practice, for $N > 100$ 
and $c \ge 3$ one can safely assume that RRGs constructed
using the method presented above will consist of a
single connected component.

While the local structure of a random network
is characterized by the degree distribution $P(k)$,
the large-scale structure is captured 
by the distribution of shortest path lengths (DSPL) 
between pairs of distinct nodes.
Properties of the DSPL, 
whose probability mass function is denoted by $P(L = \ell)$, 
have been studied in random networks with different degree distributions
\cite{Newman2001,Chung2002,Dorogovtsev2003DSPL,Fronczak2004,Hofstad2005,Blondel2007,Esker2008,Shao2009,Katzav2015,Nitzan2016,Steinbock2017,Tishby2018b,Katzav2018,Budnick2023}.
It was shown that in random networks 
that consist of a single connected component,
whose degree distribution has a finite variance,
the mean distance between pairs of distinct nodes scales like
$\langle L \rangle \sim \ln N$
\cite{Newman2001,Dorogovtsev2003DSPL,Esker2008}.
This implies that random networks are small-world networks
\cite{Chung2002,Fronczak2004}.
The DSPL of  
configuration model networks 
can be calculated using 
recursion equations
\cite{Katzav2015,Nitzan2016,Tishby2018b}.
In the case of RRGs there is a closed-form analytical expression
for the DSPL
\cite{Hofstad2005,Tishby2022},
which follows a discrete Gompertz distribution
\cite{Gompertz1825,Shklovskii2005}.

The eccentricity of a given node is the maximum distance from this node to any
other node in the network.
Thus, the eccentricity is a useful centrality measure
which quantifies how each node is situated in the global structure
of the network
\cite{Batool2014}.
A node with low eccentricity may be considered as centrally located, 
because it is reachable from
all the other nodes via relatively short paths.
In contrast, a node with high eccentricity may be considered as more peripheral.
The maximum eccentricity among all the nodes in a network is the
diameter of the network
\cite{Bollobas1981,Bollobas1982,Fernholz2007}
while the minimum eccentricity is the
radius of the network.
In the context of communication processes, the eccentricity characterizes
the worst-case scenario in the time it takes for a message to reach its target.
The computation of node eccentricities in large graphs is a non-trivial task,
particularly in small-world networks, whose short diameters render approximation
errors significant.
Recent advances include exact and scalable algorithms tailored specifically
for such networks
\cite{Li2019}.

In this paper we use methods of extreme value theory to
derive a closed-form analytical expression for the 
distribution of eccentricities (DoE) of RRGs that consist of $N$ nodes of degree $c$.
By providing the full distribution rather than a single characteristic length scale,
we present a detailed view of the large-scale structure.
In spite of the fact that the degrees of all the nodes are the same,
their eccentricities exhibit non-trivial variations. 
Using the tail-sum formula 
and the Euler-Maclaurin summation,
we obtain a closed-form expression for 
the mean eccentricity $\langle E \rangle$.
We calculate the
mode of the DoE,
which exhibits a staircase profile as a function of the network size.
Interestingly, the mode is given by
$E_{\rm mode} ={\rm Round} \left(  \langle E \rangle  \right)$,
where ${\rm Round}( x )$ is the nearest integer to $x$.
We also calculate the variance ${\rm Var}(E)$ and 
show that it exhibits oscillations as a function of the
network size $N$.

The paper is organized as follows.
In Sec. 2 we review the distribution of shortest path lengths.
In Sec. 3 we derive a closed-form expression for the distribution of eccentricities.
In Sec. 4 we calculate the mean eccentricity $\langle E \rangle$.
In Sec. 5 we calculate the mode of the distribution of eccentricities.
In Sec. 6 we calculate the variance of the distribution of eccentricities.
The results are discussed in Sec. 7 and summarized in Sec. 8.
In Appendix A we calculate the probability $P_{\rm M}(N,c)$ 
that an RRG of size $N$ and degree $c$ 
will consist of two or more components.

\section{The distribution of shortest path lengths}

The DSPL 
between pairs of random nodes
in RRGs consisting of $N \gg 1$ nodes of degree $c$ 
was calculated in Ref. 
\cite{Tishby2022}
using recursion equations.
This derivation led to a closed-form expression
for the tail of the DSPL,
which is given by
\cite{Tishby2022}

\begin{equation}
P(L> \ell) = 
\exp \left( - \frac{ e^{b \ell} - 1 }{ \beta } \right),  
\label{eq:PLtail}
\end{equation}

\noindent
where

\begin{equation}
b=\ln(c-1) 
\label{eq:b}
\end{equation}

\noindent
is the shape parameter and

\begin{equation}
\beta = \frac{(c-2)N}{c} 
\label{eq:beta}
\end{equation}

\noindent
is the scale parameter,
in agreement with Ref. 
\cite{Hofstad2005}.
Note that in this paper we use $e^{x}$ or $\exp(x)$ interchangeably to assist the
readability of the equations.
The tail distribution presented in Eq. (\ref{eq:PLtail}) 
is a discrete version of the Gompertz distribution
\cite{Gompertz1825,Shklovskii2005}.
Note that the Gompertz distribution is often expressed in terms
of the parameter $\eta = 1/\beta$,
which may be referred to as the inverse scale parameter.

For $c \ge 3$ the parameters of the Gompertz distribution satisfy $b > 0$ and $\beta > 0$.
This implies that $P(L > \ell)$ is a monotonically decreasing function, as expected from
a tail distribution.
Moreover, it exhibits a decreasing sigmoid-like shape,
or a smoothed Heaviside step function.
As $c$ is increased, the sigmoid function becomes steeper,
which implies that the probability mass function $P(L = \ell)$ becomes narrower.
As $N$ is increased, the step shifts to the right,
which implies that distances in the network become longer.
Inserting $\ell=0$ in Eq. (\ref{eq:PLtail}), we obtain
$P(L>0) = 1$, which confirms the normalization of the distribution.
The probability mass function 
is given by

\begin{equation}
P(L=\ell) = P(L > \ell - 1) - P(L > \ell).
\label{eq:PLm}
\end{equation}

\noindent
Note that for $\ell \gg 1$, 
the $-1$ term in the exponent
in Eq. (\ref{eq:PLtail}) is negligible compared to $e^{b \ell}$.
Thus
the tail distribution 
can be reduced to

\begin{equation}
P(L> \ell) \simeq 
\exp \left( - \frac{ e^{b \ell} }{ \beta } \right),
\label{eq:PLtail2}
\end{equation}

\noindent
where the $\simeq$ symbol means that the expression on the right
hand side provides a good approximation, 
which becomes exact in the limit of $N \rightarrow \infty$.
The mean distance between pairs 
of distinct nodes in an RRG is given by
\cite{Tishby2022}

\begin{equation}
\langle L \rangle \simeq  
\frac{\ln N}{\ln (c-1)} 
- \frac{ \ln c - \ln (c-2) +\gamma}{\ln (c-1)}
+ \frac{1}{2},
\label{eq:Lmean}
\end{equation}

\noindent
where $\gamma$ is the Euler-Mascheroni constant
\cite{Finch2003}. 
The variance of the DSPL was found to take the form
\cite{Tishby2022}

\begin{equation}
{\rm Var}(L) \simeq 
\frac{ \pi^2 }{6 [ \ln(c-1) ]^2 }  + \frac{1}{12}.
\label{eq:VarL}
\end{equation}

\noindent
This result implies that except for the limit of very small networks,
the variance ${\rm Var}(L)$ does not depend on the network size $N$
but only on the degree $c$.
Interestingly, the mean and variance exhibit oscillations as a function of $c$
around the curves obtained from 
Eqs. (\ref{eq:Lmean}) and (\ref{eq:VarL}), respectively
\cite{Tishby2022}.
These oscillations reflect the discrete nature of the shell structure around a random node.

\section{The distribution of eccentricities}

Below we use methods of extreme value theory 
\cite{Fisher1928,Mises1936,Gnedenko1943}
to calculate the distribution of eccentricities
in RRGs. Extreme value theory is a branch of statistics focused on the analysis of rare, 
high-impact events, which fall outside the typical range of the underlying statistical distribution. 
To illustrate these ideas, consider a random variable $X$ that follows the
distribution $P(X=x)$. Sampling $n$ independent instances of $X$, denoted by
$X_1,X_2,\dots,X_n$, the maximum $X_{\rm max}$ among these random variables follows the
cumulative distribution $P(X_{\rm max} \le x) = [P(X \le x)]^n$.
This cumulative distribution can also be expressed in terms of the tail
distribution of $X$, namely
$P(X_{\rm max} \le x) = [1 - P(X > x)]^n$.
For sufficiently large values of $n$,
the distribution $P(X_{\rm max} \le x)$ is determined
almost entirely by the tail of the underlying distribution,
where $P(X > x) \ll 1$.
This implies that $P(X_{\rm max} \le x)$ can be approximated by
$P(X_{\rm max} \le x) \simeq \exp \left[ - n P(X > x) \right]$.
Since the details of $P(X=x)$ at moderate values of $x$ are washed out,
the distribution of the maximum collapses onto one of three universal forms:
the Gumbel distribution
\cite{Gumbel1935},
for exponential-like tails of $P(X > x)$, 
the Fr\'echet distribution
\cite{Frechet1927},
for power-law tails, 
and the Weibull distribution
\cite{Weibull1951}
for tails that are truncated from above.
In particular, the Gumbel distribution is given by

\begin{equation}
P(X_{\rm max} > x) = 1 - \exp \left[ - \exp \left( - \frac{ x - \mu}{\beta} \right) \right],
\label{eq:Gumbel}
\end{equation}

\noindent
where $\mu$ is the location parameter and $\beta$ is the scale parameter.

In order to calculate the distribution of eccentricities, we consider a random
node $i$. The distances between the other $N-1$ nodes in the network and
the node $i$ follow the distribution $P(L=\ell)$.
In a sufficiently large network these distances are expected to be 
only weakly correlated, 
where the strongest correlations are between pairs of nearest neighbors.
The distance between nearest neighbors $j$ and $j'$ is $\ell_{j,j'}=1$.
This implies that the distances from $j$ and $j'$ to any other node $i$ satisfy
$|\ell_{i,j} - \ell_{i,j'}| \le 1$.
However, the fraction of pairs of nodes that are nearest neighbors is of the
order of $1/N$, which implies that these correlations constitute a subleading
effect of order ${\mathcal O}(1/N)$.

Using the framework of extreme value theory presented above,
and assuming that the distances $\ell_{i,j}$
between a random node $i$ and all the other nodes in the network are 
only weakly correlated,
the distribution of eccentricities can be
expressed in the form

\begin{equation}
P(E \le \ell) \simeq \left[ P(L \le \ell) \right]^{N-1},   
\label{eq:PEell}
\end{equation}

\noindent
where $P(L \le \ell)$ is the cumulative distribution of the DSPL.
Using the fact that $P(L \le \ell) = 1 - P(L > \ell)$
and replacing $N-1$ by $N$,
we obtain

\begin{equation}
P(E \le \ell) \simeq \left[ 1 - P(L > \ell) \right]^{N}.
\label{eq:PEell2}
\end{equation}

\noindent
For sufficiently large values of $\ell$, the tail distribution satisfies
$N \left[ P(L > \ell) \right]^2 \ll 1$. 
In this regime, one can exponentiate the term in
the square brackets and obtain

\begin{equation}
P(E \le \ell) \simeq \exp \left[ -   N  P(L > \ell)  \right].
\label{eq:PEell3}
\end{equation}

\noindent
Thus, the tail distribution of eccentricities is given by

\begin{equation}
P(E > \ell) \simeq 1 - \exp \left[ -   N  P(L > \ell)  \right].
\label{eq:PEell3p}
\end{equation}

\noindent
Inserting $P(L > \ell)$ from Eq. (\ref{eq:PLtail2}) into Eq. (\ref{eq:PEell3p}),
we obtain

\begin{equation}
P(E > \ell) \simeq 1 - \exp \left[ -    N \exp \left( - \frac{ e^{b \ell}  }{\beta}  \right) \right].
\label{eq:PEell4}
\end{equation}

\noindent
Expressing $N$ in the form $\exp(\ln N)$ 
and rearranging terms,
it is found that the DoE of an RRG that consists of $N$ nodes of degree $c$
is given by

\begin{equation}
P(E>\ell) \simeq 1 - \exp \left[ - \exp  \left( - \frac{ e^{b \ell} - \mu }{\beta} \right) \right],
\label{eq:PEell7}
\end{equation}

\noindent
where the shape parameter $b$ is  
the same as in Eq. (\ref{eq:b}),
the scale parameter 
$\beta$ is the same as in Eq. (\ref{eq:beta}) 
and

\begin{equation}
\mu =   \frac{ c-2 }{c} N  \ln N  
\label{eq:mu}
\end{equation}

\noindent
is the location parameter.
Throughout the paper we refer to $P(E>\ell)$, given by Eq. (\ref{eq:PEell7}),
as the tail distribution of the DoE.
The probability mass function of the DoE
is given by

\begin{equation}
P(E=\ell) = P(E > \ell - 1) - P(E > \ell),
\label{eq:PEm}
\end{equation}

\noindent
where $P(E > \ell)$ is given by Eq. (\ref{eq:PEell7}).
 
Eq. (\ref{eq:PEell7}) resembles the form of a Gumbel distribution,  
given by Eq. (\ref{eq:Gumbel}).
In fact, this is not a Gumbel distribution, because the linear term $x$ 
in the Gumbel distribution is replaced by the exponential term $(c-1)^{\ell}$.
However, it can be shown that
after appropriate centering and rescaling of the variable $\ell$,
the resulting distribution converges to a Gumbel distribution
in the infinite network limit.
This is in agreement with the
Fisher–Tippett–Gnedenko theorem, often called the extreme value theorem
\cite{Fisher1928,Mises1936,Gnedenko1943}.

\begin{figure}
\centerline{
\includegraphics[width=6.5cm]{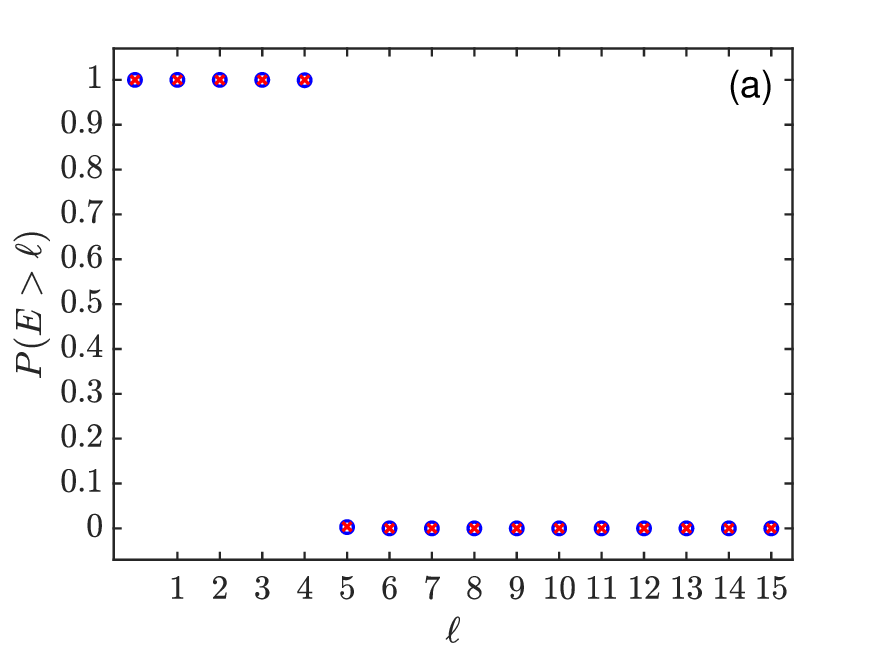}
\includegraphics[width=6.5cm]{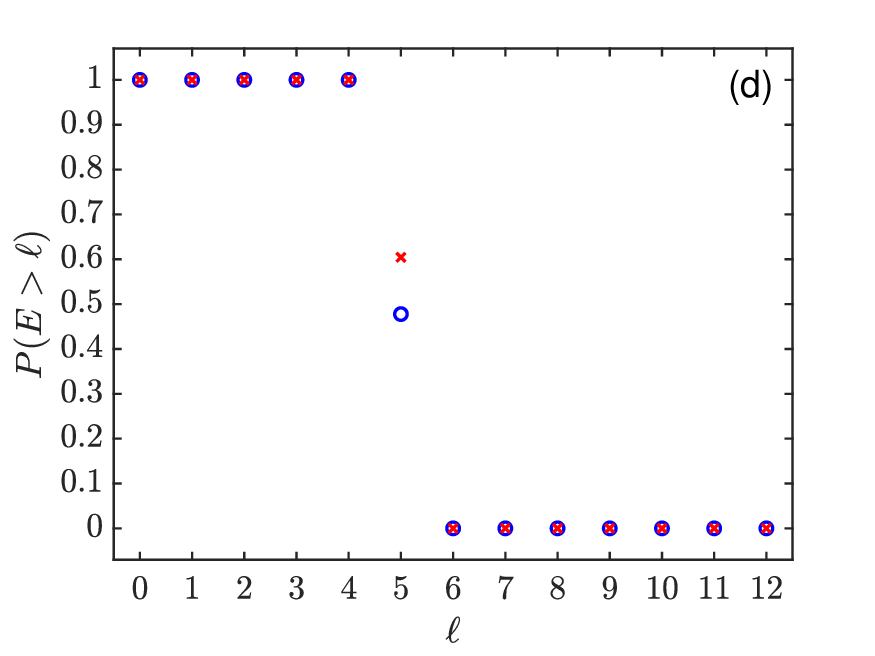}
}
\centerline{
\includegraphics[width=6.5cm]{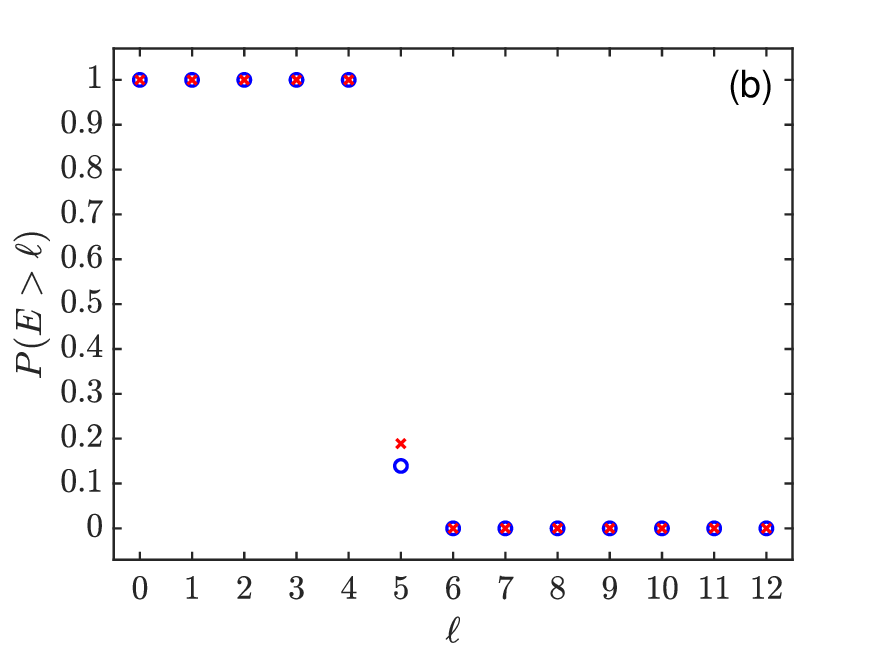}
\includegraphics[width=6.5cm]{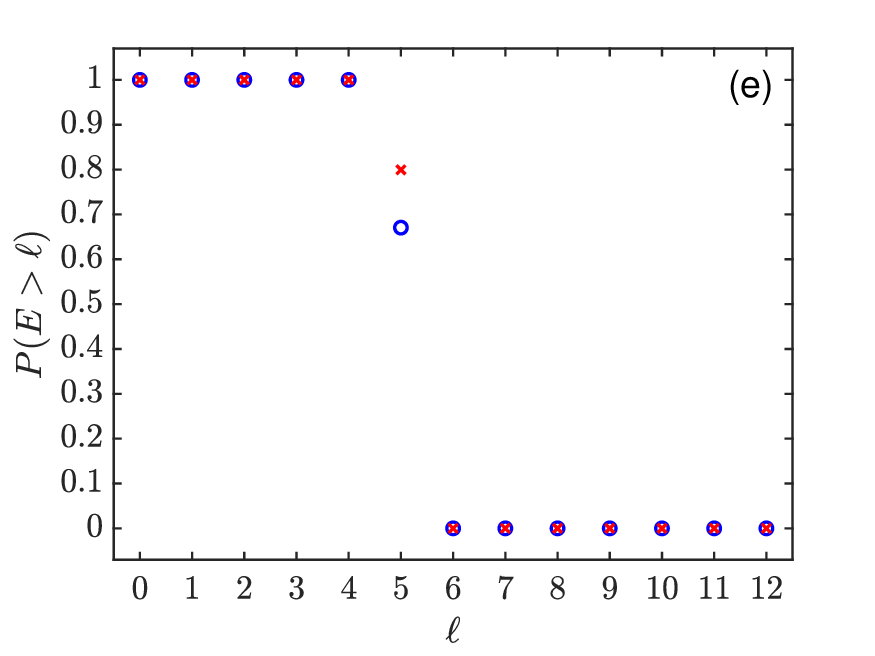}
}
\centerline{
\includegraphics[width=6.5cm]{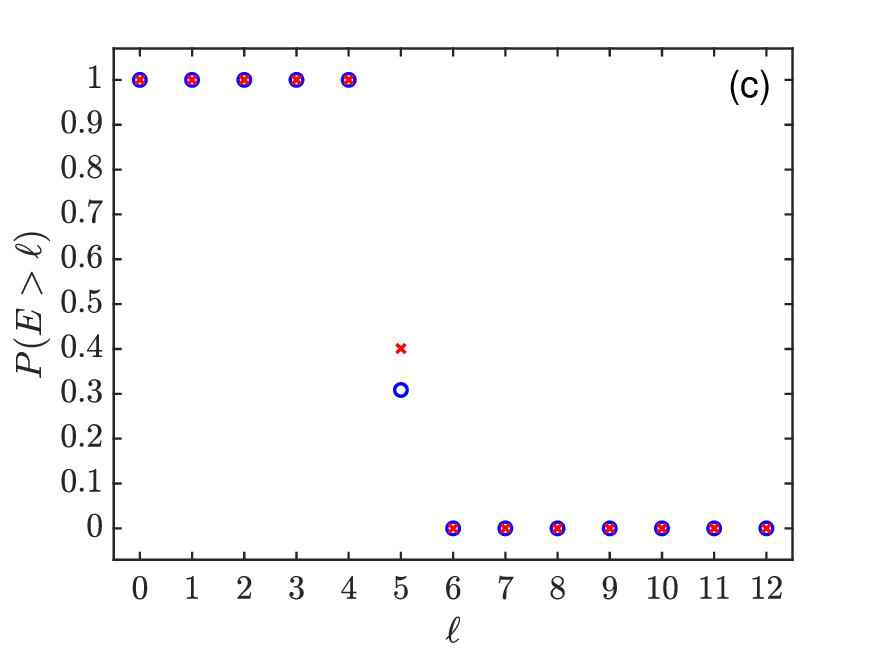}
\includegraphics[width=6.5cm]{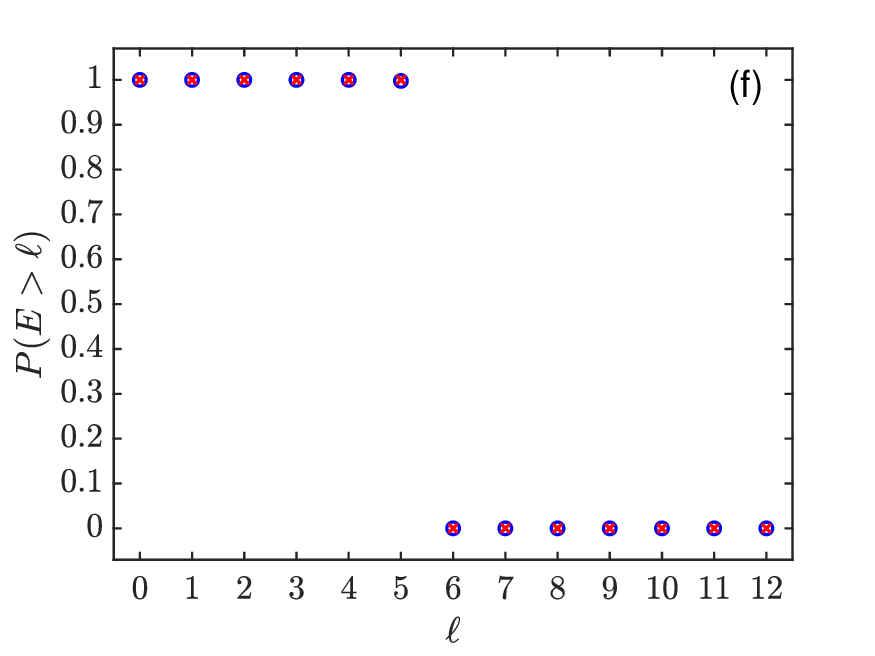}
}
\caption{
Analytical results  ($\times$)
for the distribution of eccentricities
$P(E > \ell)$
of nodes in RRGs that consist of nodes of degree $c=5$ 
and network sizes of (a) $N=160$;
(b) $N=242$; (c) $N=272$;
(d) $N=296$; (e) $N=322$ and
(f) $N=440$.
The analytical results, 
obtained from Eq. (\ref{eq:PEell7}),
are in reasonable agreement with the results obtained 
from computer simulations ($\circ$),
where the deviations are due to finite-size effects.
The simulation results were averaged over 1000 network instances.
For $N=160$ the eccentricity of essentially all the nodes is
$E=5$. As $N$ is increased, nodes of eccentricity $E=6$ emerge
and their weight increases until at $N=440$ essentially
all the nodes are of eccentricity $E=6$.
}
\label{fig:1}
\end{figure}

In Fig. \ref{fig:1} we present
analytical results ($\times$)
for the tail distribution of eccentricities
$P(E > \ell)$ in RRGs 
that consist of nodes of degree $c=5$ and network
sizes of (a) $N=160$; (b) $N=242$; (c) $N=272$;
(d) $N=296$; (e) $N=322$ and (f) $N=440$.
The analytical results,
obtained from Eq. (\ref{eq:PEell7}),
are in reasonable agreement with the results obtained 
from computer simulations ($\circ$),
where the discrepancies are due to finite-size effects.
For $N=160$ the eccentricity of essentially all the nodes is
$E=5$. As $N$ is increased, nodes of eccentricity $E=6$ emerge
and their weight increases until at $N=440$ essentially
all the nodes are of eccentricity $E=6$.
The simulation results presented in Fig. \ref{fig:1} and in all the other Figures
in this paper are based on $1,000$ network instances for 
each value of the network size $N$.
In these simulations the distances were calculated using the 
breadth-first search algorithm.

\begin{figure}
\centerline{
\includegraphics[width=6.5cm]{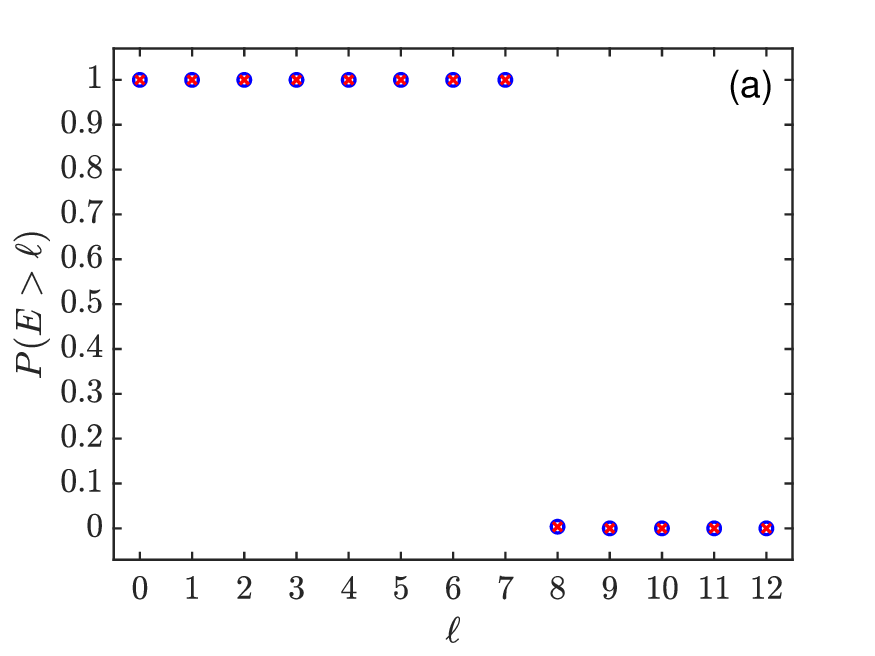}
\includegraphics[width=6.5cm]{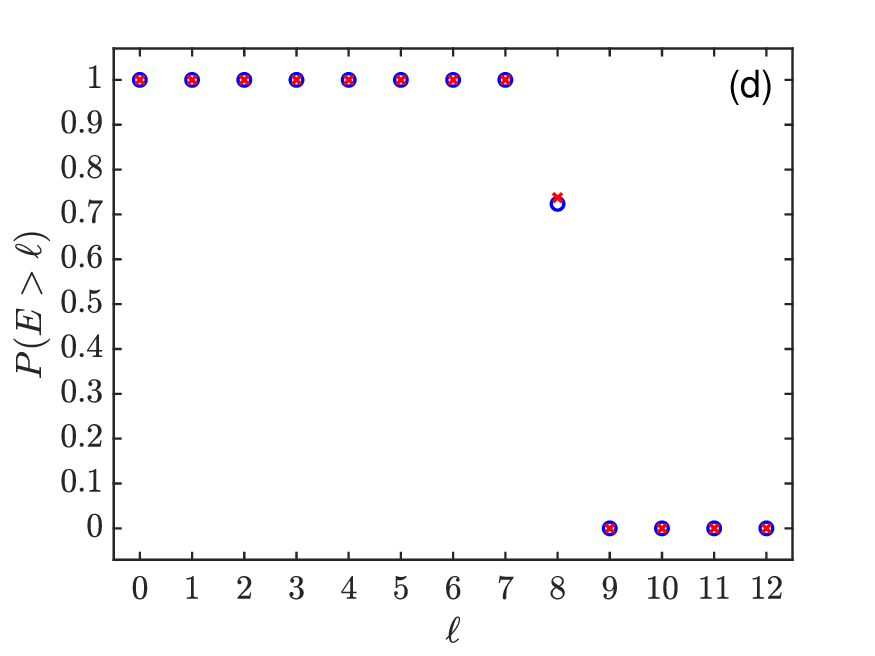}
}
\centerline{
\includegraphics[width=6.5cm]{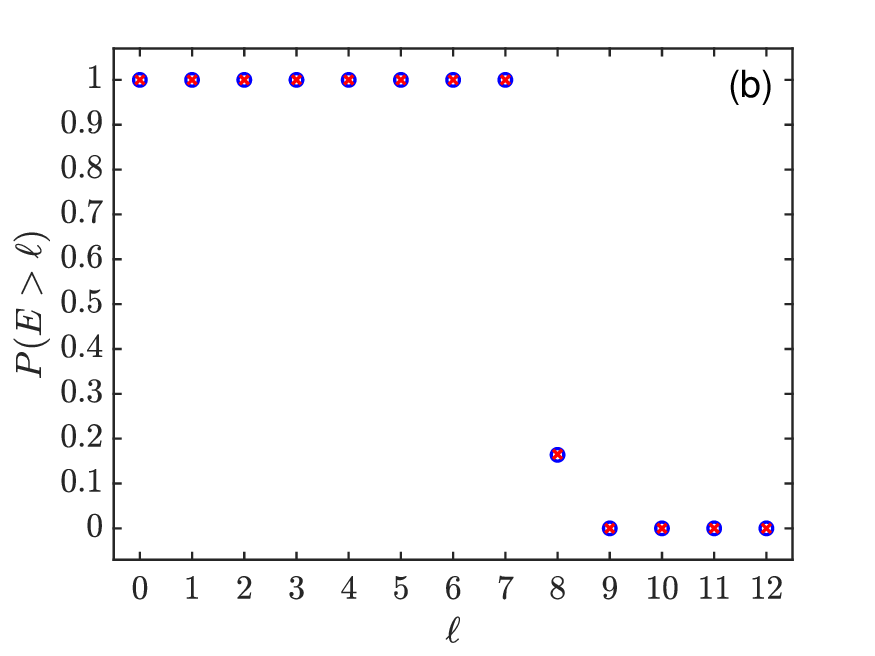}
\includegraphics[width=6.5cm]{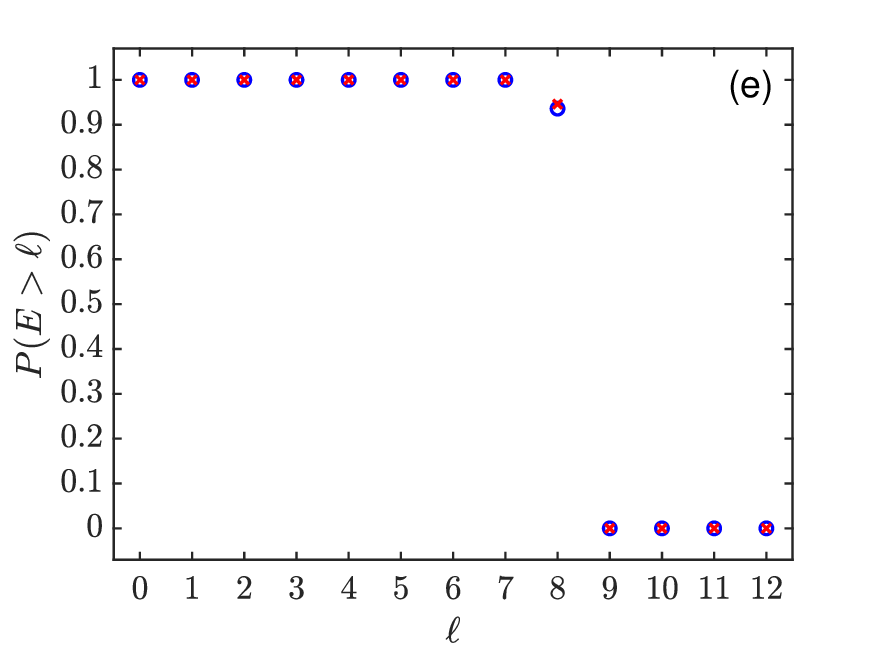}
}
\centerline{
\includegraphics[width=6.5cm]{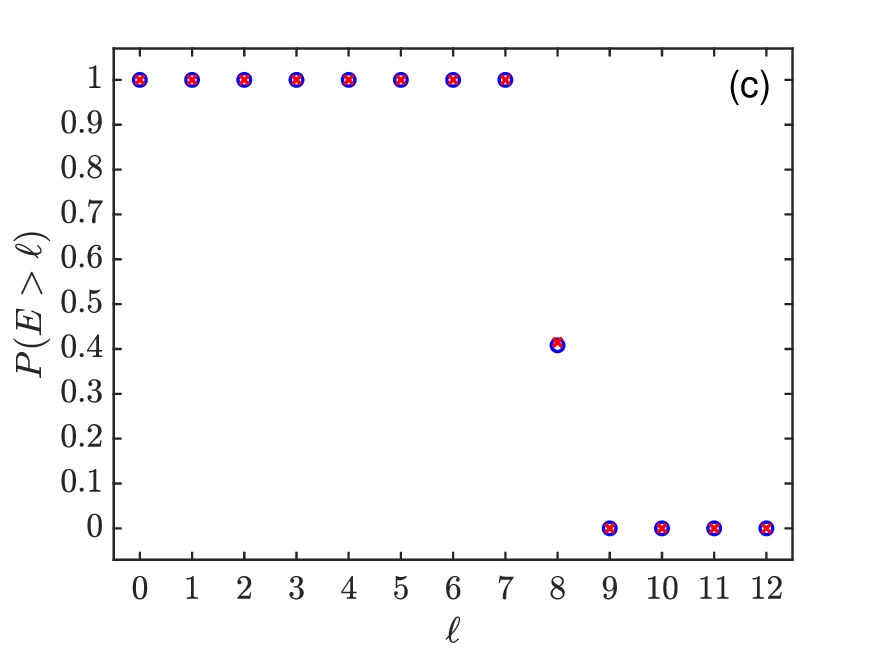}
\includegraphics[width=6.5cm]{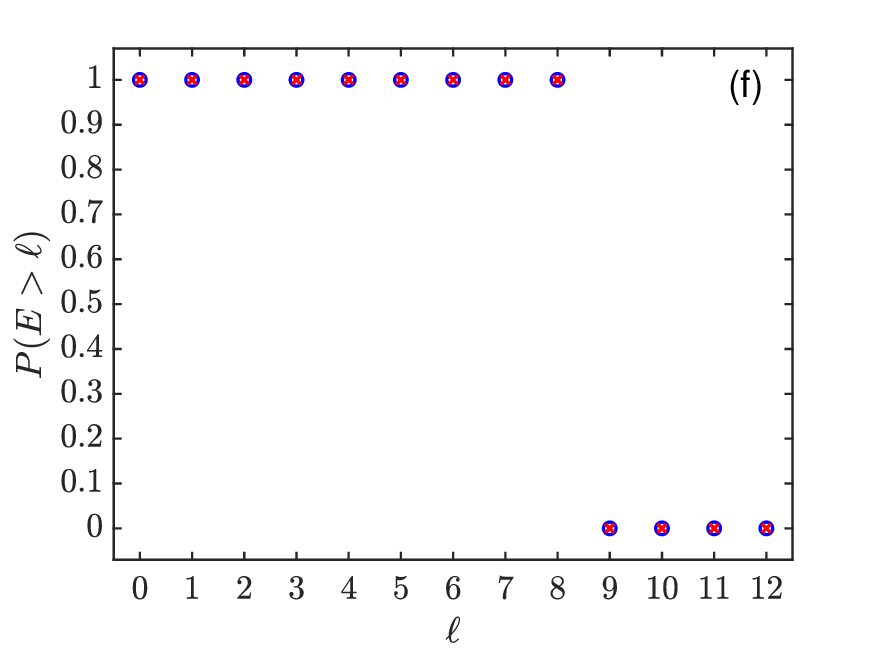}
}
\caption{
Analytical results  ($\times$)
for the distribution of eccentricities
$P(E > \ell)$
of nodes in RRGs that consist of nodes of degree $c=5$ 
and network sizes of (a) $N=7,500$;
(b) $N=10,000$; (c) $N=11,000$;
(d) $N=12,000$; (e) $N=13,000$ and
(f) $N=20,000$.
The analytical results, 
obtained from Eq. (\ref{eq:PEell7}),
are in very good agreement with the results obtained 
from computer simulations ($\circ$).
The simulation results were averaged over 1000 network instances.
For $N=7,500$ the eccentricity of essentially all the nodes is
$E=8$. As $N$ is increased, nodes of eccentricity $E=9$ emerge
and their weight increases until at $N=20,000$ essentially
all the nodes are of eccentricity $E=9$.
}
\label{fig:2}
\end{figure}

In Fig. \ref{fig:2} we present
analytical results ($\times$)
for the tail distribution of eccentricities
$P(E > \ell)$ in RRGs 
that consist of nodes of degree $c=5$ and network
sizes of (a) $N=7,500$; (b) $N=10,000$; (c) $N=11,000$;
(d) $N=12,000$; (e) $N=13,000$ and (f) $N=20,000$.
The analytical results,
obtained from Eq. (\ref{eq:PEell7}),
are in very good agreement with the results obtained 
from computer simulations ($\circ$).
For $N=7,500$ the eccentricity of essentially all the nodes is
$E=8$. As $N$ is increased, nodes of eccentricity $E=9$ emerge
and their weight increases until at $N=20,000$ essentially
all the nodes are of eccentricity $E=9$.

\begin{figure}
\centerline{
\includegraphics[width=6.5cm]{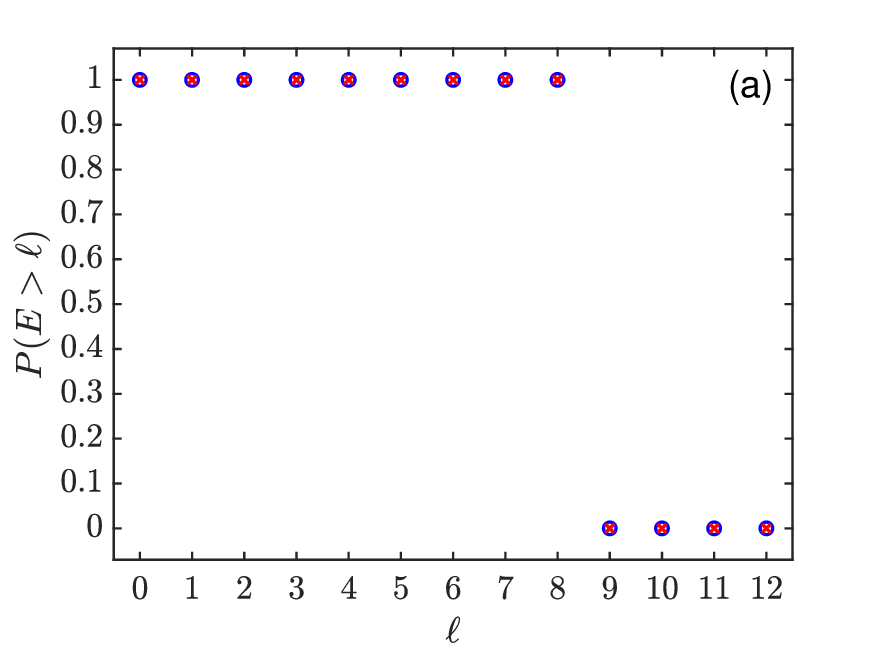}
\includegraphics[width=6.5cm]{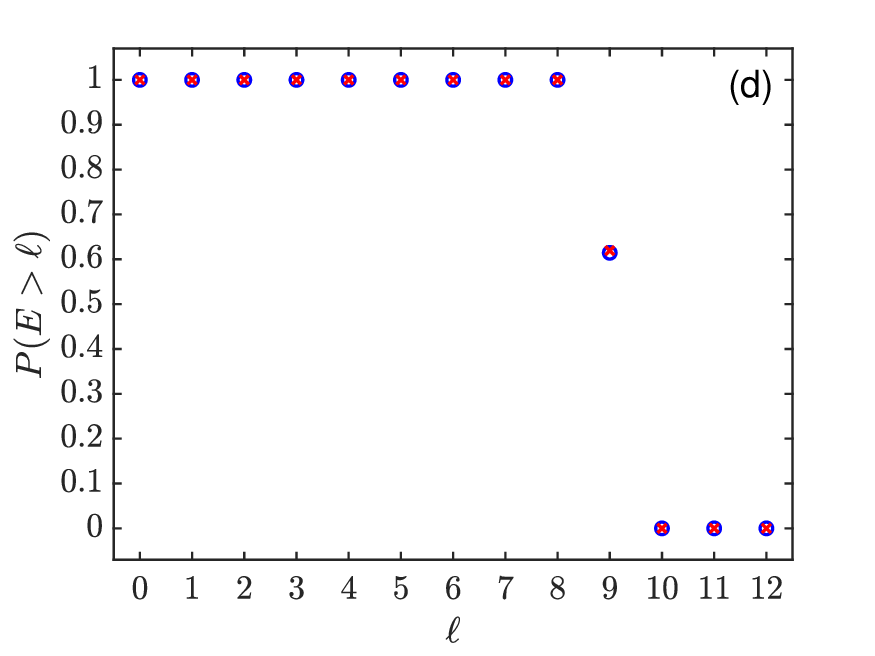}
}
\centerline{
\includegraphics[width=6.5cm]{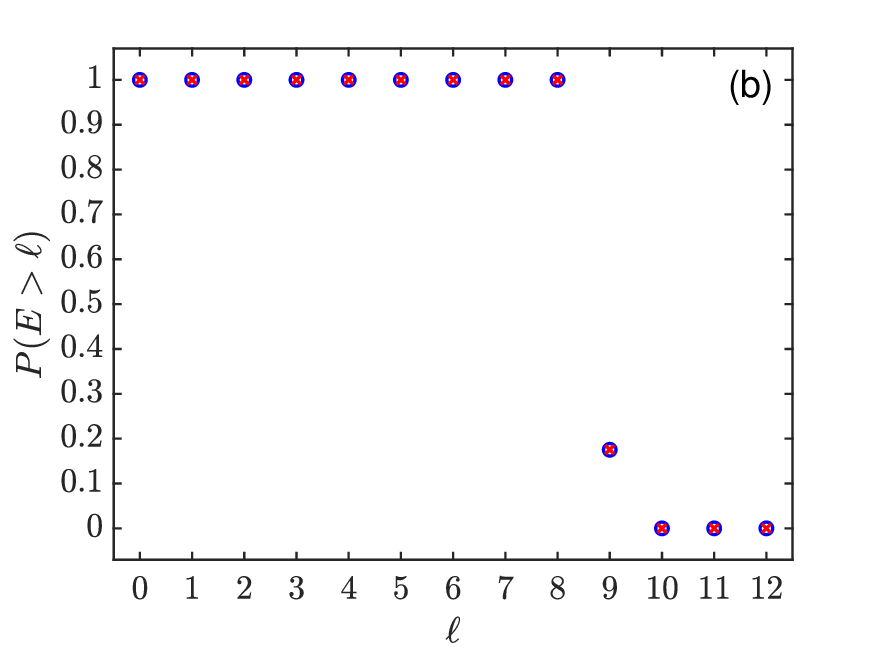}
\includegraphics[width=6.5cm]{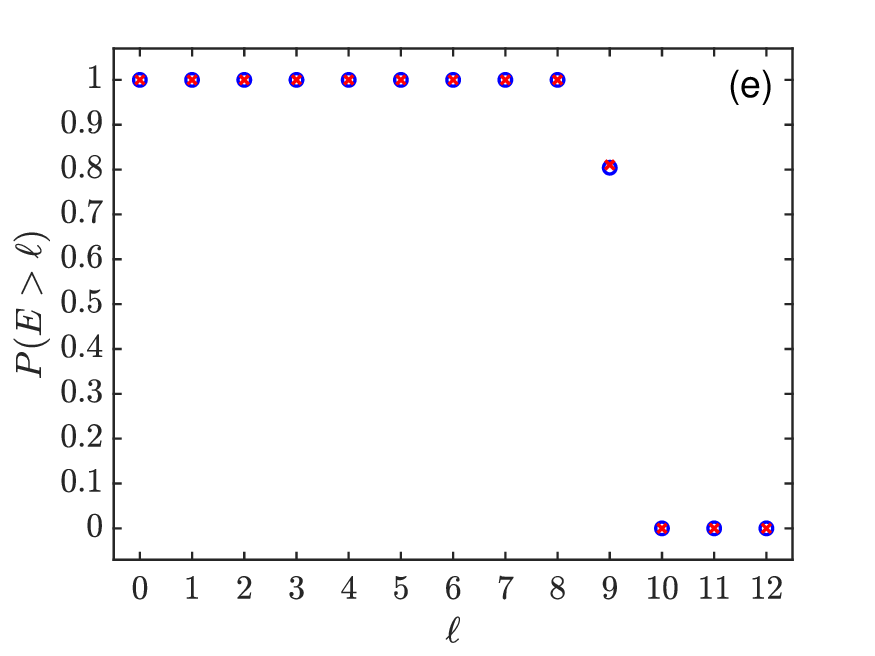}
}
\centerline{
\includegraphics[width=6.5cm]{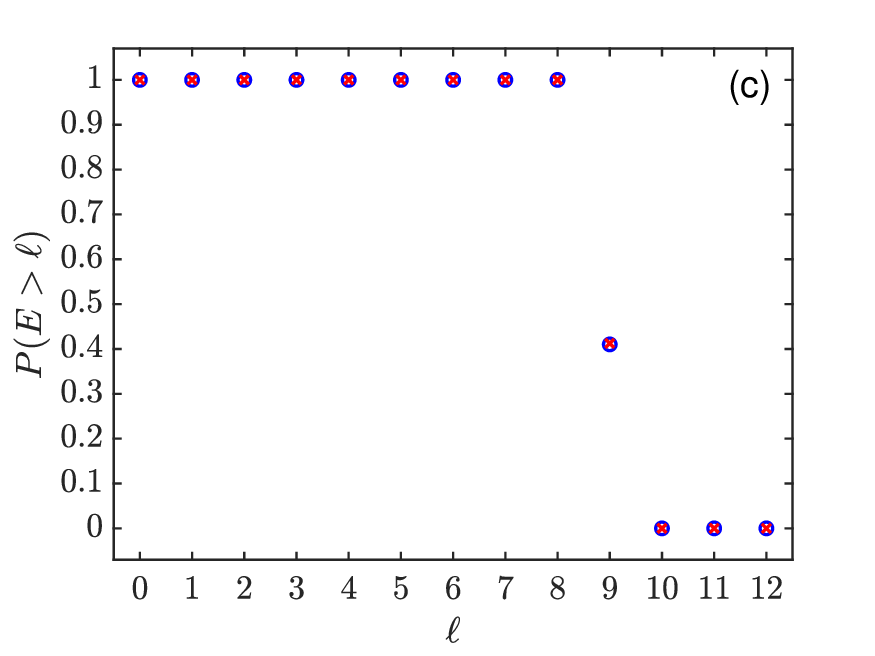}
\includegraphics[width=6.5cm]{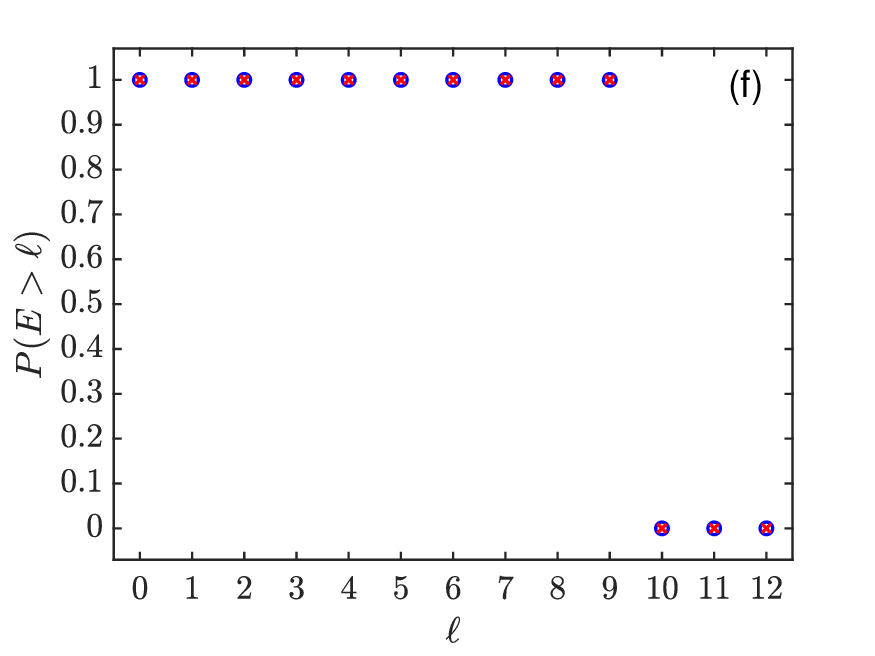}
}
\caption{
Analytical results  ($\times$)
for the distribution of eccentricities
$P(E > \ell)$
of nodes in RRGs that consist of nodes of degree $c=5$ 
and network sizes of (a) $N=22,000$;
(b) $N=36,000$; (c) $N=39,000$;
(d) $N=41,000$; (e) $N=43,000$ and
(f) $N=55,000$.
The analytical results, 
obtained from Eq. (\ref{eq:PEell7}),
are in excellent agreement with the results obtained 
from computer simulations ($\circ$).
The simulation results were averaged over 1000 network instances.
For $N=22,000$ the eccentricity of essentially all the nodes is
$E=9$. As $N$ is increased, nodes of eccentricity $E=10$ emerge
and their weight increases until at $N=55,000$ essentially
all the nodes are of eccentricity $E=10$.
}
\label{fig:3}
\end{figure}

In Fig. \ref{fig:3} we present
analytical results ($\times$)
for the tail distribution of eccentricities
$P(E > \ell)$ in RRGs 
that consist of nodes of degree $c=5$ and network
sizes of (a) $N=22,000$; (b) $N=36,000$; (c) $N=39,000$;
(d) $N=41,000$; (e) $N=43,000$ and (f) $N=55,000$.
The analytical results,
obtained from Eq. (\ref{eq:PEell7}),
are in excellent agreement with the results obtained 
from computer simulations ($\circ$).
For $N=22,000$ the eccentricity of essentially all the nodes is
$E=9$. As $N$ is increased, nodes of eccentricity $E=10$ emerge
and their weight increases until at $N=55,000$ essentially
all the nodes are of eccentricity $E=10$.

Figs. \ref{fig:1}-\ref{fig:3} demonstrate that the analytical results capture
the behavior of the DoE of RRGs of various sizes. In general, the comparison
between the theory and the simulations improves dramatically as the network
size $N$ is increased.
In fact, in Figs. \ref{fig:1}-\ref{fig:3} the discrepancy between the theory and the simulation
results for $P(E>\ell)$ appears only in a single value of $\ell$,
which is the transition point between the plateau at $1$ on the left
and the plateau at $0$ on the right.
To provide a more quantitative assessment of the convergence,
we focus below on the transition values of $\ell$
for different ranges of network sizes.

\begin{figure}
\centerline{
\includegraphics[width=7.5cm]{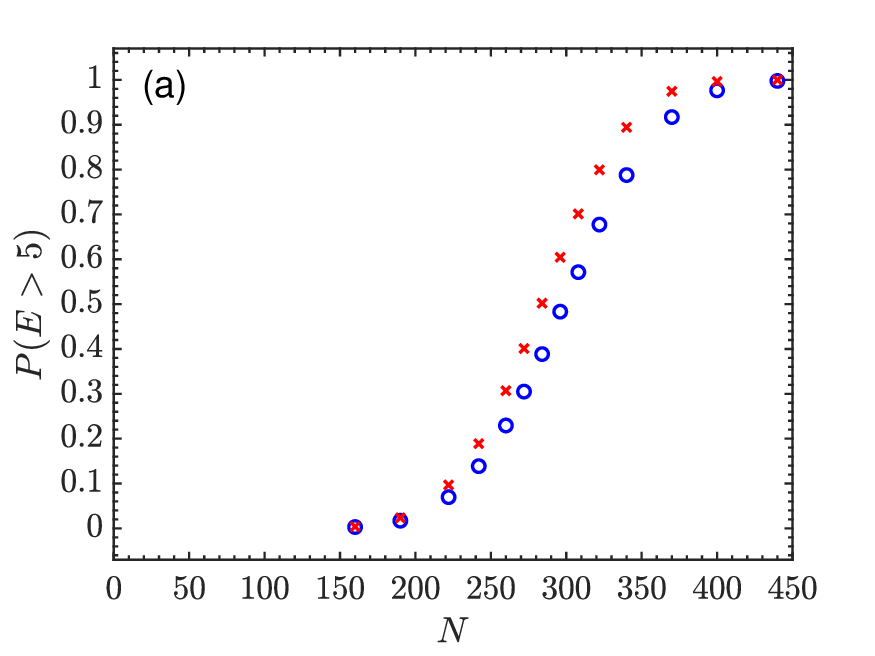}
\includegraphics[width=7.5cm]{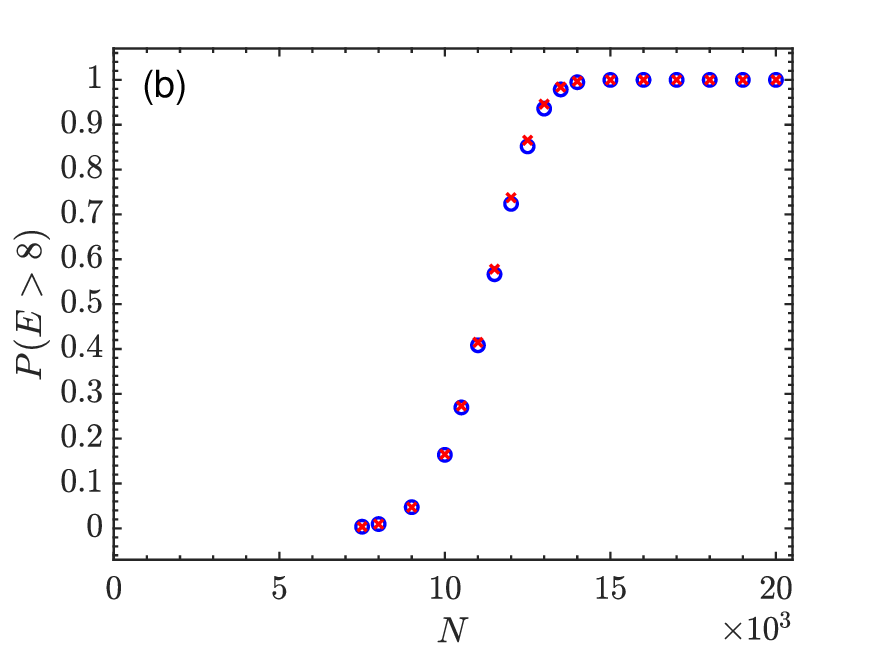}
}
\centerline{
\includegraphics[width=7.5cm]{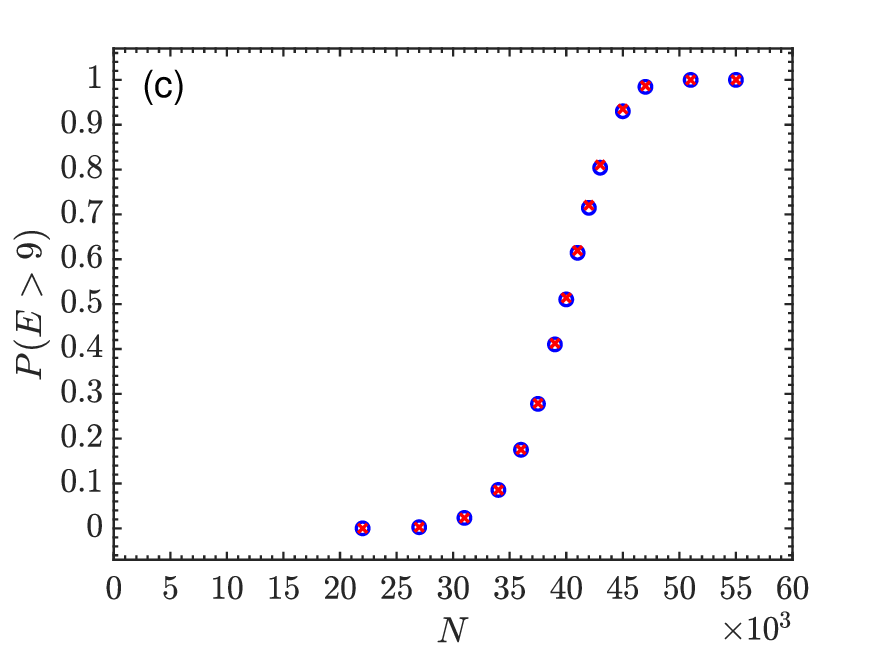}
}
\caption{
Analytical results ($\times$) for the probability 
(a) $P(E>5)$; 
(b) $P(E>8)$
and
(c) $P(E>9)$,
as a function of the 
network size $N$ for RRGs consisting of nodes of degree $c=5$,
obtained from Eq. (\ref{eq:PEell7}).
The analytical results are found to be systematically larger than the
corresponding simulation results ($\circ$).
For each value of $\ell$, the probability $P(E>\ell)$ is shown for the
window of network sizes in which it changes from zero to one.
The discrepancy between the theory and simulation results 
decreases as the network size is increased.
}
\label{fig:4}
\end{figure}

In Fig. \ref{fig:4} we present
analytical results ($\times$) for the probabilities
(a) $P(E>5)$; 
(b) $P(E>8)$
and
(c) $P(E>9)$,
as a function of the 
network size $N$ for RRGs consisting of nodes of degree $c=5$,
obtained from Eq. (\ref{eq:PEell7}).
The analytical results are systematically larger than the
corresponding simulation results ($\circ$).
For each value of $\ell$, the probability $P(E>\ell)$ is shown for the
range of network sizes over which it increases from zero to one.
The discrepancy between the theory and simulation results 
decreases as the network size is increased,
confirming that it is a result of finite-size effects.

\section{The mean eccentricity}

The mean eccentricity $\langle E \rangle$
of nodes in an RRG consisting of $N$ nodes of degree $c$
can be calculated using the tail-sum formula 
\cite{Pitman1993}

\begin{equation}
\langle E \rangle = \sum_{\ell=0}^{N-2} P(E > \ell),
\label{eq:Emean1}
\end{equation}

\noindent
where $P(E>\ell)$ is given by Eq. (\ref{eq:PEell7}).
Since $P(E>\ell)$ is vanishingly small for 
$\ell \ge N-1$, Eq. (\ref{eq:Emean1}) can be replaced by 

\begin{equation}
\langle E \rangle = \sum_{\ell=0}^{\infty} P(E > \ell).
\label{eq:Emean2}
\end{equation}

\noindent
Using the Euler-Maclaurin formula, we obtain
\cite{Apostol1999,Bender1999}

\begin{equation}
\langle E \rangle 
\simeq 
\frac{1}{2} + \int_{0}^{\infty} P(E > \ell) d \ell.
\label{eq:Emean3}
\end{equation}

\noindent
Note that the integral approximation,
presented in Eq. (\ref{eq:Emean3}),
ignores the discrete 
nature of the shell structure around a random node.
It is thus expected to 
describe the smooth trend of the mean eccentricity.
Inserting $P(E>\ell)$ from Eq. (\ref{eq:PEell7}) into Eq. (\ref{eq:Emean3}), 
we obtain

\begin{equation}
\langle E \rangle 
\simeq 
\frac{1}{2} + \int_{0}^{\infty}  
\left\{ 1 - \exp \left( - \exp  \left[ - \frac{ (c-1)^{\ell} - \mu }{\beta} \right] \right) \right\}
d \ell,
\label{eq:Emean4}
\end{equation}

\noindent
In order to carry out the integration, we express Eq. (\ref{eq:Emean4}) in the form

\begin{equation}
\langle E \rangle 
\simeq 
\frac{1}{2} + \int_{0}^{\infty}  
\left\{ 1 -  \exp \left( - N \exp \left[   -  \frac{  (c-1)^{\ell}  }{\beta}   \right] \right) \right\} 
d \ell.
\label{eq:Emean5}
\end{equation}
 
\noindent
Using the change of variables 

\begin{equation}
y = (c-1)^{\ell}, 
\end{equation}

\noindent
we obtain

\begin{equation}
\langle E \rangle 
\simeq 
\frac{1}{2} + 
\frac{1}{\ln (c-1)}
\int_{1}^{\infty}  
\left\{  1 -  \exp \left[ - N \exp  \left( -   \frac{ y }{\beta}    \right) \right]   \right\}
\frac{ d y }{y},
\label{eq:Emean5b}
\end{equation}
 
\noindent
Further changing the integration variable to 

\begin{equation}
x =  \frac{y}{\beta},
\end{equation}

\noindent
we obtain

\begin{equation}
\langle E \rangle 
\simeq 
\frac{1}{2} + 
\frac{1}{\ln (c-1)}
\int_{1/\beta}^{\infty}  
\left\{  1 -  \exp \left[ - N \exp  \left( -  x    \right) \right]  \right\}
\frac{ d x }{x},
\label{eq:Emean6}
\end{equation}

\noindent
Expressing the numerator on the right hand side of Eq. (\ref{eq:Emean6})
in the form

\begin{equation}
1 -  \exp \left[ - N \exp  \left( -  x    \right) \right] 
=
\int_{0}^{N e^{-x}} e^{-u} du,
\label{eq:identity}
\end{equation}

\noindent
we obtain

\begin{equation}
\langle E \rangle 
\simeq
\frac{1}{2} + 
\frac{1}{\ln (c-1)}
\int_{1/\beta}^{\infty}  
\frac{d x}{x}  
\int_{0}^{N e^{-x}} e^{-u} du.
\label{eq:Emean6p}
\end{equation}

\noindent
Exchanging the order of the integrals, we obtain

\begin{equation}
\langle E \rangle 
\simeq
\frac{1}{2} + 
\frac{1}{\ln (c-1)}
\int_{0}^{N e^{-1/\beta}} e^{-u} d u
\int_{1/\beta}^{\ln (N/u)}  
\frac{d x}{x}. 
\label{eq:Emean7}
\end{equation}

\noindent
The inner integral is given by

\begin{equation}
\int_{1/\beta}^{\ln (N/u)}  
\frac{d x}{x}  
=
\ln \left[ \ln \left( \frac{N}{u} \right) \right] + \ln \beta.
\label{eq:inner}
\end{equation}

\noindent
Inserting the solution of the inner integral from Eq. (\ref{eq:inner})
into Eq. (\ref{eq:Emean7}), we obtain

\begin{equation}
\langle E \rangle 
\simeq
\frac{1}{2} + 
\frac{1}{\ln (c-1)}
\left\{
\int_{0}^{N e^{-1/\beta}} e^{-u}
\ln \left[ \ln \left( \frac{N}{u} \right) \right]
d u
+
\ln \beta \int_{0}^{N e^{-1/\beta}} e^{-u} du
\right\}.
\label{eq:Emean8}
\end{equation}

\noindent
For sufficiently large $N$, the upper limits of the integrals can be extended 
to infinity, with exponentially small error. 
Carrying out the second integral then yields

\begin{equation}
\langle E \rangle 
\simeq 
\frac{1}{2} + 
\frac{1}{\ln (c-1)}
\int_{0}^{\infty} e^{-u}
\ln \left[ \ln \left( \frac{N}{u} \right) \right]
d u
+
\frac{ \ln \beta }{ \ln (c-1) }.
\label{eq:Emean9}
\end{equation}

\noindent
In order to carry out the first integration, we express the logarithmic
function in the integrand in the form

\begin{equation}
\ln \left[ \ln \left( \frac{N}{u} \right) \right]
=
\ln \ln N + \ln \left( 1 - \frac{ \ln u }{ \ln N } \right).
\label{eq:LnLn}
\end{equation}

\noindent
Using the expansion

\begin{equation}
\ln (1-z) = -z - \frac{z^2}{2} 
- \dots,
\end{equation}

\noindent
where $z = \ln u / \ln N$,
we obtain

\begin{equation}
\ln \left[ \ln \left( \frac{N}{u} \right) \right]
=
\ln \ln N 
- \frac{\ln u}{\ln N}
+ {\mathcal O}  \left[ \left( \frac{ \ln u }{ \ln N } \right)^2 \right].
\label{eq:LnLn2}
\end{equation}

\noindent
We then insert the results back into Eq. (\ref{eq:Emean9}) and carry out the
integration term by term.
The integral over the first term on the right hand side of Eq. (\ref{eq:LnLn2})
simply yields $\ln \ln N$, while the second term yields

\begin{equation}
\int_{0}^{\infty} e^{-u}
\frac{\ln u}{\ln N}
d u
=
- \frac{\gamma}{\ln N},
\label{eq:lnu}
\end{equation}

\noindent
and the third term is ${\mathcal O} \left[ \left( 1 / \ln N \right)^2 \right]$.
Putting all these results together,
we obtain

\begin{equation}
\langle E \rangle = \frac{1}{2} + 
\frac{1}{\ln (c-1)}
\left(
\ln \beta +
\ln \ln N 
+ \frac{\gamma}{\ln N}
\right)
+ 
{\mathcal O} \left[      \left( \frac{1}{   \ln N  }     \right)^2   \right].
\label{eq:Emean10}
\end{equation}

\noindent
Writing Eq. (\ref{eq:Emean10}) explicitly in terms of $N$ and $c$
and rearranging terms, we obtain

\begin{eqnarray}
\langle E \rangle   &=&  
\frac{\ln N}{\ln (c-1)}
+
\frac{ \ln \ln N }{\ln(c-1)}
- \frac{ \ln c - \ln  (c-2) }{\ln(c-1)} 
+ \frac{1}{2}
\nonumber \\
&+& 
\frac{ \gamma }{\ln (c-1) \ln N}
+ {\mathcal O}  \left[   \left(   \frac{1}{\ln N}      \right)^2 \right].
\label{eq:Emean12}
\end{eqnarray}

\noindent
In the large network limit, it is sufficient to use a more compact expression,
which is given by

\begin{eqnarray}
\langle E \rangle    = 
\frac{\ln N}{\ln (c-1)}
+
\frac{ \ln \ln N }{\ln(c-1)}
- \frac{ \ln c - \ln  (c-2) }{\ln(c-1)} 
+ \frac{1}{2}
+ {\mathcal O}     \left(   \frac{1}{\ln N}      \right).
\label{eq:Emean12b}
\end{eqnarray}

For sufficiently large networks,
a direct numerical evaluation of the
mean eccentricity $\langle E \rangle$
can be done using 
an approximate form of the 
tail-sum formula, given by Eq. (\ref{eq:Emean2}).
To this end, we introduce the effective distance

\begin{equation}
\ell^{*} = \bigg\lfloor 
\frac{ \ln N }{ \ln (c-1) } + \frac{ \ln \ln N }{ \ln (c-1) } + \frac{1}{2}
\bigg\rfloor,
\label{eq:x}
\end{equation}

\noindent
where $\lfloor x \rfloor$ is the integer part of $x$,
which provides a crude approximation for $\langle E \rangle$.
Using this notation we obtain a much better approximation for the mean eccentricity,
which is given by

\begin{equation}
\langle E \rangle  \simeq  \ell^{*} - 1  + P(E>\ell^{*}-1) + P(E>\ell^{*}) + P(E>\ell^{*}+1),
\label{eq:Edense}
\end{equation}
 
\noindent
where the term $\ell^{*} - 1$ on the right hand side of Eq. (\ref{eq:Edense})
accounts for the sum over the probabilities $P(E>\ell)$ for
$\ell=0,1,\dots,\ell^{*}-2$, which can be
approximated by $P(E>\ell) \simeq 1$. 
The next three terms account for the
narrow range of eccentricities in which the tail distribution decreases 
sharply, 
while the probabilities $P(E>\ell)$ for $\ell \ge \ell^{*}+2$
are negligible.
This expression is particularly useful for numerical evaluation
of the mean eccentricity.

\begin{figure}
\centerline{
\includegraphics[width=8cm]{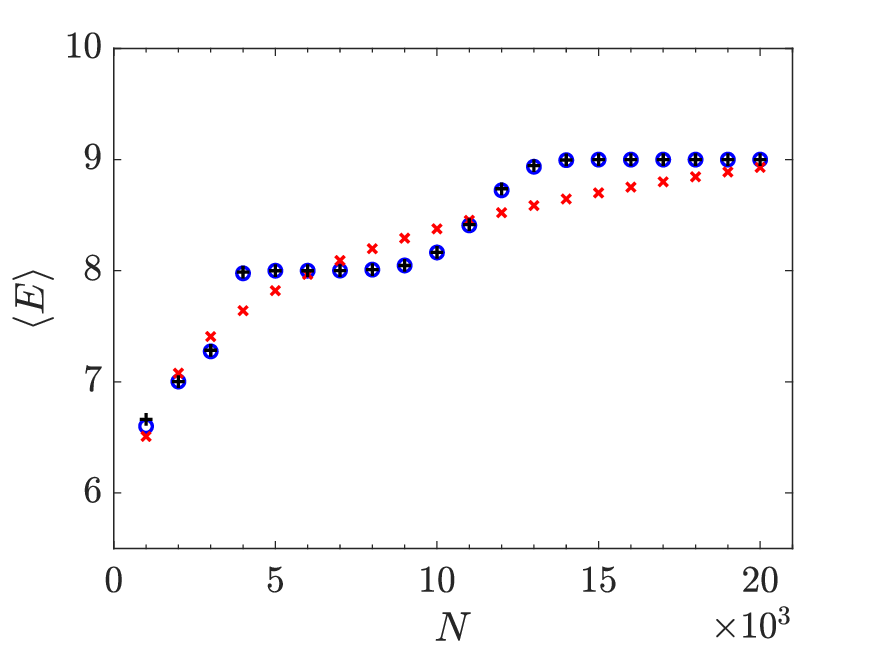}
}
\caption{
Analytical results ($\times$) 
and simulation results ($\circ$)
for the mean eccentricity $\langle E \rangle$ 
in RRGs that consist of nodes of degree $c=5$, as a function of the
network size $N$.
While the simulation results tend to lock in to integer values,
the analytical results,
obtained from Eq. (\ref{eq:Emean12b}),
follow a smoothed-out contour.
We also present  approximate results ($+$), obtained from Eq. (\ref{eq:Edense}).
These results, which are based on an approximate form of the tail-sum formula,
are in very good agreement with the simulation results.
However, the analytical expression given by Eq. (\ref{eq:Emean12b}) 
elucidates the scaling of $\langle E \rangle$
as a function of $N$ and $c$ in a more transparent way.
}
\label{fig:5}
\end{figure}

In Fig. \ref{fig:5} we present 
analytical results ($\times$) 
and simulation results ($\circ$)
for the mean eccentricity $\langle E \rangle$ 
in RRGs that consist of nodes of degree $c=5$, as a function of the
network size $N$.
While the simulation results tend to lock in to integer values,
the analytical results,
obtained from Eq. (\ref{eq:Emean12b}),
follow a smoothed-out contour.
Note that the difference between the more precise expression given by Eq. 
(\ref{eq:Emean12}) and the more compact expression given by Eq. (\ref{eq:Emean12b})
is less than one percent for all the data points presented in Fig. \ref{fig:5}.
Thus, using Eq. (\ref{eq:Emean12}) would not improve the agreement with the simulation results.
We also present  approximate results ($+$), obtained from Eq. (\ref{eq:Edense}).
These results, which are based on an approximate form of the tail-sum formula,
are in very good agreement with the simulation results.
However, the analytical expression given by Eq. (\ref{eq:Emean12b}) 
elucidates the scaling of $\langle E \rangle$
as a function of $N$ and $c$ in a more transparent way.

\section{The mode of the distribution of eccentricities}

The mode of the DoE in an RRG that consists 
of $N$ nodes of degree $c \ge 3$ is the
most probable eccentricity of nodes in 
the network.
For sufficiently large networks, the DoE is essentially a binary distribution,
namely for a given choice of $N \gg 1$ and $c \ge 3$,
there is some value of $\ell$ such that the only significant
contributions to the probability mass function are from 
$P(E=\ell)$ and $P(E=\ell+1)$.
Thus, in case that $P(E=\ell) > P(E=\ell+1)$ the mode is
$E_{\rm mode}=\ell$, while in the opposite case it is 
$E_{\rm mode}=\ell+1$.
Plotting the mode $E_{\rm mode}$ as a function of $N$
for a given value of $c$, one obtains a step function,
where $E_{\rm mode}$ takes only integer values.
A similar phenomenon was observed in ER networks
under the condition that they are sufficiently dense
\cite{Chung2001}.

The location of the step edge (as a function of $N$) between 
$E_{\rm mode}=\ell$ 
and 
$E_{\rm mode}=\ell+1$
is determined by
$P(E=\ell) = P(E=\ell+1)$.
Expressing the DoE in terms of the tail distribution $P(E > \ell)$, 
one finds that the step edge is determined 
by the condition $P(E > \ell) = 1/2$.
Inserting $P(E>\ell) = 1/2$ into the left-hand side of Eq. (\ref{eq:PEell7})
and rearranging terms, we obtain

\begin{equation}
\frac{ e^{b \ell} - \mu }{\beta} = \ln \left( \frac{1}{\ln 2} \right).
\label{eq:ebell1}
\end{equation}

\noindent
Writing Eq. (\ref{eq:ebell1}) explicitly in terms of $N$ and $c$,
we obtain

\begin{equation}
 \frac{ N }{ \ln 2 } \ln \left( \frac{N}{ \ln 2 } \right) 
=
\frac{ c (c-1)^{\ell} }{    (c-2) \ln 2 }.
\label{eq:NlnN}
\end{equation}

\noindent
Extracting the network size $N$ from Eq. (\ref{eq:NlnN}),
it is found that the location of the step edge between 
$E_{\rm mode}=\ell$ and $E_{\rm mode}=\ell+1$
is given by

\begin{equation}
N_{\ell,\ell+1}(c) = \frac{ c (c-1)^{\ell} }{   (c-2) W_0 \left[ \frac{ c (c-1)^{\ell} }{    (c-2) \ln 2    } \right] },
\label{eq:Ncell}
\end{equation}

\noindent
where $W_0(x)$ is the Lambert W function
\cite{Olver2010}.
In other words, $N_{\ell,\ell+1}(c)$ is the network size below which the
mode of the DoE is $\ell$ and above which the mode is $\ell+1$.
Thus, the width of the step in which $E_{\rm mode} = \ell$ is 

\begin{equation}
\Delta_{\ell}(c) = N_{\ell,\ell+1}(c) - N_{\ell-1,\ell}(c).
\label{eq:Wcell}
\end{equation}

\noindent
Extracting the eccentricity $\ell$ in terms
of $N$ and $c$ from Eq. (\ref{eq:NlnN}),
it is found that

\begin{equation}
\ell = \frac{ \ln \left[   \frac{ (c-2) N }{c} \ln \left( \frac{N}{\ln 2 } \right)  \right] }{\ln (c-1)}.
\label{eq:ell}
\end{equation}

\noindent
This implies that for a given choice of $N$ and $c$, 
the mode of the DoE is given by

\begin{equation}
E_{\rm mode} = \left\lceil \frac{ \ln N }{ \ln (c-1) } + \frac{ \ln \ln \left( \frac{N}{ \ln 2 } \right) }{\ln (c-1)}
- \frac{ \ln c - \ln (c-2) }{\ln (c-1)} \right\rceil,
\label{eq:Lmode1}
\end{equation}

\noindent
where $\lceil x \rceil$ is the smallest integer that is greater than or equal to $x$.
Expanding the term $\ln \ln (N/ \ln 2 )$ in powers of $1/\ln N$, 
we obtain

\begin{equation}
E_{\rm mode} 
= 
\left\lceil \frac{ \ln N }{ \ln (c-1) } 
+ \frac{ \ln \ln  N }{\ln (c-1)}
- \frac{ \ln c - \ln (c-2) }{\ln (c-1)} 
\right\rceil,
\label{eq:Emode2}
\end{equation}

\noindent
where the error in the expression inside the   
brackets 
on the right hand side of Eq. (\ref{eq:Emode2})
is of order ${\mathcal O}(1 / \ln N)$.
Using the identity

\begin{equation}
\lceil x \rceil = {\rm Round} \left( x + \frac{1}{2} \right),
\end{equation}

\noindent
one observes that

\begin{equation}
E_{\rm mode} = {\rm Round} \left( \langle E \rangle \right),
\end{equation}

\noindent
where $\langle E \rangle$ is given by Eq. (\ref{eq:Emean12b}).
This implies that as the network size $N$ is varied,
the mode $E_{\rm mode}$ forms a staircase function that closely follows the smooth curve of the
mean eccentricity $\langle E \rangle$.

\begin{figure}
\centerline{
\includegraphics[width=8cm]{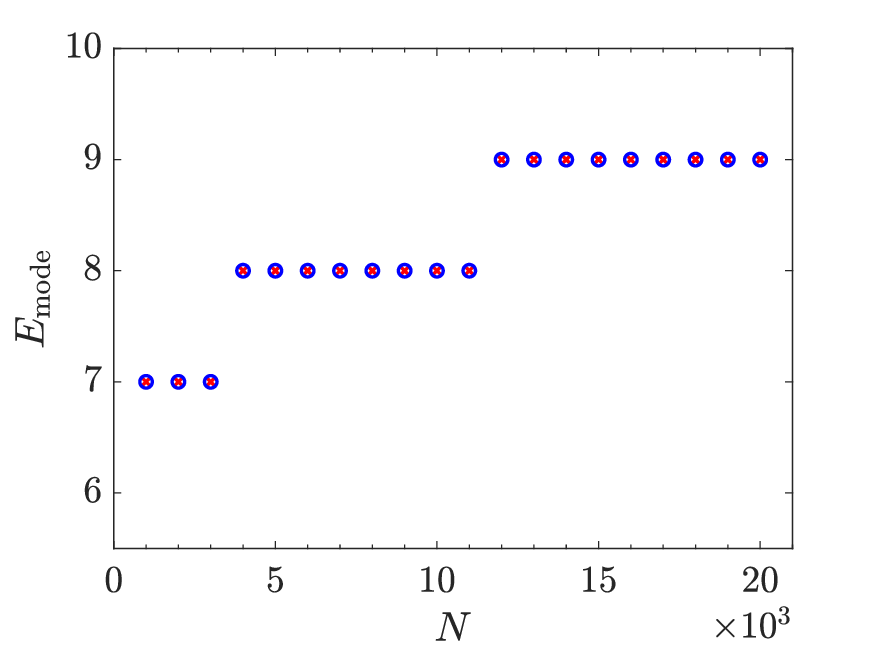}
}
\caption{
Analytical results ($\times$) for  
the mode $E_{\rm mode}$
of the DoE
in RRGs that consist of nodes of degree $c=5$,
as a function of the network size $N$.
The analytical results,
obtained from 
Eq. (\ref{eq:Emode2}),
are in very good agreement with the results obtained from computer simulations
($\circ$).
The mode $E_{\rm mode}$ exhibits a non-decreasing staircase function 
of the network size $N$.
The width of the steps increases as $N$ is increased. 
}
\label{fig:6}
\end{figure}

In Fig. \ref{fig:6} we present
analytical results ($\times$) for  
the mode $E_{\rm mode}$
of the DoE
in RRGs that consist of nodes of degree $c=5$,
as a function of the network size $N$.
The analytical results,
obtained from 
Eq. (\ref{eq:Emode2}),
are in very good agreement with the results obtained from computer simulations
($\circ$).
The mode $E_{\rm mode}$ exhibits a non-decreasing staircase function 
of the network size $N$.
The width of the steps increases as $N$ is increased.

\section{The variance of the distribution of eccentricities}

Using the tail-sum formula, the second moment of the DoE
can be expressed in the form
\cite{Pitman1993}

\begin{equation}
\langle E^2 \rangle = \sum_{\ell=0}^{N-2}
(2\ell+1) P(E>\ell).
\label{eq:VarE1}
\end{equation}

\noindent
Since the probability $P(E>\ell)$, given by Eq. (\ref{eq:PEell7}),
is vanishingly small for $\ell \ge N-1$, the upper limit of the summation can be
changed from $N-2$ to $\infty$, without any noticeable 
change in the result.
Therefore,

\begin{equation}
\langle E^2 \rangle =   \sum_{\ell=0}^{\infty}
(2\ell+1) P(E>\ell).
\label{eq:VarE2}
\end{equation}

\noindent
Using the Euler-Maclaurin formula, we obtain

\begin{equation}
\langle E^2 \rangle 
\simeq
\frac{1}{4} +
\int_{0}^{\infty} (2\ell+1) P(E>\ell) d \ell.
\label{eq:VarE3}
\end{equation}

\noindent
Inserting $P(E>\ell)$ from Eq. (\ref{eq:PEell7}) into Eq. (\ref{eq:VarE3}),
we obtain

\begin{equation}
\langle E^2 \rangle 
\simeq
\frac{1}{4} +
\int_{0}^{\infty} (2\ell+1) 
\left\{ 1 - \exp \left[ - N \exp \left( - \frac{ (c-1)^{\ell} }{\beta} \right) \right] \right\}
d \ell.
\label{eq:VarE3p1}
\end{equation}

\noindent
To simplify the integrand, we introduce the variable

\begin{equation}
y = (c-1)^{\ell}.
\label{eq:yell}
\end{equation}

\noindent
Expressing the distance $\ell$ in terms of $y$,
we obtain

\begin{equation}
\ell = \frac{ \ln y }{ \ln (c-1) },
\label{eq:elly}
\end{equation}

\noindent
which implies that the first term in the integral on the right hand side
of Eq. (\ref{eq:VarE3p1}) can be expressed in the form

\begin{equation}
2 \ell + 1 = 2 \frac{ \ln y }{ \ln (c-1) } + 1.
\label{eq:2ell}
\end{equation}

\noindent
Differentiating both sides of Eq. (\ref{eq:elly}), 
we obtain

\begin{equation}
d \ell = \frac{ d y }{ y \ln (c-1) }.
\label{eq:dell}
\end{equation}

\noindent
Expressing $\ell$ on the right hand side of Eq. (\ref{eq:VarE3p1}) in terms of $y$,
we obtain

\begin{equation}
\langle E^2 \rangle 
\simeq
\frac{1}{4} + \frac{1}{ \ln (c-1) }
\int_{1}^{\infty} \left[ 2 \frac{ \ln y }{ \ln (c-1) } + 1 \right] 
\left\{ 1 - \exp \left[ - N \exp \left( - \frac{y}{\beta} \right) \right] \right\}
\frac{ d y }{y}.
\end{equation}

\noindent
Replacing the integration variable $y$ by $x = y/\beta$,
we obtain

\begin{equation}
\langle E^2 \rangle 
\simeq
\frac{1}{4} + 
\frac{1}{ \ln (c-1) } \int_{1/\beta}^{\infty} 
\left[ 2 \frac{ \ln (\beta x) }{ \ln (c-1) } + 1 \right] 
\left[ 1 - \exp \left( - N e^{-x} \right) \right] \frac{ d x }{x}.
\end{equation}

\noindent
Using Eq. (\ref{eq:identity}),
we obtain

\begin{equation}
\langle E^2 \rangle 
\simeq
\frac{1}{4} + 
\frac{1}{ \ln (c-1) } \int_{1/\beta}^{\infty} 
\left[ 2 \frac{ \ln (\beta x) }{ \ln (c-1) } + 1 \right] \frac{d x}{x}
\int_{0}^{ N e^{-x} } e^{-u} d u.
\end{equation}

\noindent
Exchanging the order of integrations,
we obtain

\begin{equation}
\langle E^2 \rangle 
\simeq
\frac{1}{4} + 
\frac{1}{ \ln (c-1) } 
\int_{0}^{ N e^{-1/\beta} } e^{-u} 
J(u) d u 
\label{eq:E2ux}
\end{equation}

\noindent
where

\begin{equation}
J(u) =
\int_{1/\beta}^{\ln(N/u)} 
\left[ 2 \frac{ \ln (\beta x) }{ \ln (c-1) } + 1 \right] \frac{d x}{x}.
\label{eq:Ju}
\end{equation}

\noindent
Carrying out the integration on the right hand side of Eq. (\ref{eq:Ju}),
we obtain

\begin{equation}
J(u) = \frac{ \left\{ \ln \left[ \beta \ln \left( \frac{N}{u} \right) \right] \right\}^2 }{ \ln (c-1) }
+ \ln \left[ \beta \ln \left( \frac{N}{u} \right) \right].
\label{eq:Ju2}
\end{equation}

\noindent
Inserting $J(u)$ from Eq. (\ref{eq:Ju2}) into Eq. (\ref{eq:E2ux}),
we obtain

\begin{eqnarray}
\langle E^2 \rangle &\simeq& 
\frac{1}{4} 
+ \frac{ 1 }{ \left[ \ln (c-1) \right]^2 } 
\int_{0}^{\infty} e^{-u} 
\left\{ \ln \left[ \beta \ln \left( \frac{N}{u} \right) \right] \right\}^2 d u
\nonumber \\
&+&
\frac{1}{ \ln (c-1) } \int_{0}^{\infty} e^{-u} \ln \left[ \beta \ln \left( \frac{N}{u} \right) \right] d u.
\label{eq:E2}
\end{eqnarray}

\noindent
The second integral on the right hand side of Eq. (\ref{eq:E2}) is similar to the
one that appears in Eq. (\ref{eq:Emean9}), which yields

\begin{equation}
\int_{0}^{\infty} e^{-u} \ln \left[ \beta \ln \left( \frac{N}{u} \right) \right] d u
=
\ln \ln N + \ln \beta + \frac{ \gamma }{ \ln N } + {\mathcal O} \left[   \left(  \frac{1}{\ln N}   \right)^2     \right].
\label{eq:Int1}
\end{equation}

\noindent
Similarly, the first integral on the right hand side of Eq. (\ref{eq:E2}) yields

\begin{eqnarray}
\hspace{-0.4in}
\int_{0}^{\infty} e^{-u} 
\left\{ \ln \left[ \beta \ln \left( \frac{N}{u} \right) \right] \right\}^2 d u
&=& 
\left( \ln \ln N + \ln \beta \right)^2  
\nonumber \\
&+& 
\frac{2 \gamma \left( \ln \ln N + \ln \beta \right) }{ \ln N } 
+ {\mathcal O} \left[   \left(  \frac{1}{\ln N}   \right)^2     \right].
\label{eq:Int2}
\end{eqnarray}

\noindent
Inserting the right hand sides of Eqs. (\ref{eq:Int1}) and (\ref{eq:Int2})
into the right hand side of Eq. (\ref{eq:E2}) and rearranging terms,
we obtain

\begin{equation}
\langle E^2 \rangle =     
\langle E \rangle^2 +    {\mathcal O}    \left( \frac{\ln \ln N}{\ln N}    \right).         
\label{eq:VarE4}
\end{equation}

\noindent
This implies that the variance  

\begin{equation}
{\rm Var}(E) = \langle E^2 \rangle -   \langle E \rangle^2 
\label{eq:Var1}
\end{equation}

\noindent
satisfies

\begin{equation}
{\rm Var}(E) =  
{\mathcal O}    \left( \frac{\ln \ln N}{\ln N}    \right). 
\label{eq:Var2}
\end{equation}

\noindent
This result implies that in the continuum approximation 
the variance vanishes asymptotically as $N \rightarrow \infty$.
However, the DoE is actually a discrete distribution, which may have a non-vanishing
variance.

To calculate the actual variance of the DoE we use
the tail sum formula (\ref{eq:VarE2}), which 
can be approximated by

\begin{equation}
\langle E^2 \rangle \simeq (\ell^{*} - 1)^2 + \sum_{\ell=\ell^{*}-1}^{\ell^{*}+1} (2 \ell + 1) P(E>\ell),
\label{eq:L2dense}
\end{equation}

\noindent
where $\ell^{*}$ is given by Eq. (\ref{eq:x}).
Inserting $\langle E^2 \rangle$ from Eq. (\ref{eq:L2dense})
and $\langle E \rangle$ from Eq. (\ref{eq:Edense}) into Eq. (\ref{eq:Var1})
and rearranging terms, we obtain

\begin{eqnarray}
{\rm Var}(E) & \simeq 
P(E>\ell^{*}-1) + 3 P(E>\ell^{*}) + 5 P(E>\ell^{*}+1) 
\nonumber \\
&-  \left[ P(E>\ell^{*}-1) + P(E>\ell^{*}) + P(E>\ell^{*}+1) \right]^2.
\label{eq:VarLarge3}
\end{eqnarray}

\begin{figure}
\centerline{
\includegraphics[width=8cm]{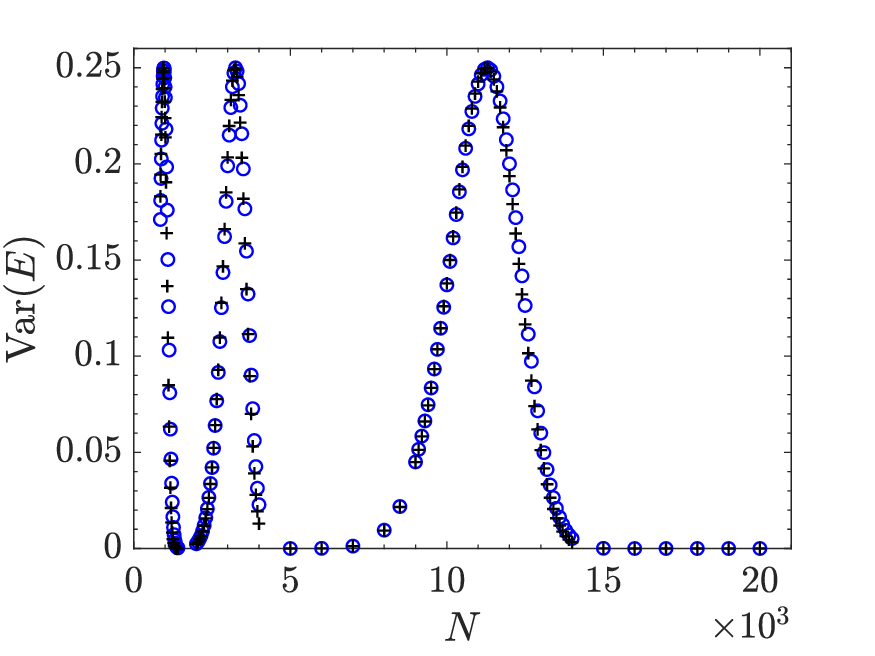}
}
\caption{
Analytical results ($+$) for the variance ${\rm Var}(E)$ of the DoE  
in RRGs that consist of nodes of degree $c=5$, as a function of the
network size $N$.
The analytical results, obtained from Eq. (\ref{eq:VarLarge3}),
are in very good agreement
with the results obtained from computer simulations ($\circ$).
The variance exhibits oscillations as a function of $N$,
which are due to the discrete nature of the eccentricity.
The peaks of these oscillations are located near the step edges,
where the mode of the DoE shifts from $\ell$ to $\ell+1$.
Since for sufficiently large networks the DoE is essentially a binary distribution,
at the peaks
$P(E=\ell) = P(E=\ell+1) \simeq 1/2$,
leading to
${\rm Var}(E) \simeq 1/4$.
}
\label{fig:7}
\end{figure}

In Fig. \ref{fig:7} we present   
analytical results ($+$) for the variance ${\rm Var}(E)$ of the DoE  
in RRGs that consist of nodes of degree $c=5$, as a function of the
network size $N$.
The analytical results, obtained from Eq. (\ref{eq:VarLarge3}),
are in very good agreement
with the results obtained from computer simulations ($\circ$).
The variance exhibits oscillations as a function of $N$,
which are due to the discrete nature of the eccentricity.
The peaks of these oscillations are located near the step edges,
where the mode of the DoE shifts from $\ell$ to $\ell+1$.
Since for sufficiently large networks the DoE is essentially a binary distribution,
at the peaks
$P(E=\ell) = P(E=\ell+1) \simeq 1/2$,
leading to
${\rm Var}(E) \simeq 1/4$.

\section{Discussion}

The eccentricity $\ell_i$ of node $i$ is an important property that captures how far the
node is from the most distant region of the network. 
It provides information about how the node is situated in the global structure of the network.
It thus complements the degree $k_i$ that is characteristic of the local 
neighborhood of the node.
It is found that even in the case of RRGs, in which all the nodes are
of degree $k_i=c$, there are variations between the eccentricities of different nodes.

The DoE provides detailed statistical information on
the large-scale structure of the network.
It is an extreme value statistic of the DSPL, which provides
a magnified view of information that is captured in the far tail of the DSPL.
The mean eccentricity $\langle E \rangle$ provides the typical distance
from a node to its farthest node, while the variance ${\rm Var}(E)$ 
quantifies the width of the distribution.
For each instance of an RRG, consisting of $N$ nodes of degree $c$,
the DoE is reduced to an eccentricity sequence of the form
$\ell_i$, $i=1,2,\dots,N$.
The largest value $\max_i \ell_i$ is referred to as the diameter of the
network, while the smallest value $\min_i \ell_i$ is referred to as the
radius of the network.
Since the DoE is a very narrow distribution, the difference between the
diameter and radius of RRGs is of order $1$.

In the context of communication on networks, the
eccentricity is important because it captures the 
worst-case communication cost from a given node. 
More specifically, it tells us how far a message, signal, or piece of information 
must travel in order to reach every other node in the network.
It is thus a key metric in the placement of servers and routers 
in a communication network.

In general, the edge-swapping process we use in the construction of the
RRGs may introduce a bias in the sampling of the networks
\cite{Klein2012}.
A possible way to avoid such bias is to reject any network instance that
includes multiple edges or self-loops. 
However, this approach comes with a possible computational cost because the fraction
of network instances that turn out to be simple graphs is small.
In fact, this fraction is given by
\cite{Janson2009,Janson2014}

\begin{equation}
P_{\rm simple}(c) = e^{ - \frac{c^2 - 1}{4} } + {\mathcal O} \left(  N^{-1} \right),
\end{equation}

\noindent
which decreases rapidly as
the degree $c$ is increased.
To confirm that the rewiring does not cause any noticeable bias in the results
presented in this paper, we repeated some of the simulations using the 
rejection approach, and compared the results with those obtained with
the rewiring approach.
We did not find any noticeable difference, confirming that the rewiring
does not bias the results for the eccentricities of RRGs.

It would be interesting to generalize the analysis presented in this paper
to a broader class of configuration model networks with non-degenerate
degree distributions $P(k)$.
A major obstacle is that unlike the case of the RRG, for more general configuration
model networks we do not have a closed-form expression for the DSPL.
In general configuration model networks the distribution $P(E=\ell)$ is expected to be broader
than in the case of RRGs.
This is due to the fact that low-degree nodes are expected to have
larger eccentricities and high-degree nodes are expected to have smaller
eccentricities.
Thus, the results presented above
suggest that degree heterogeneity may further broaden the
distribution of eccentricities.

While the diameter of RRGs was studied before
\cite{Bollobas1982},
it would also be interesting to study the distribution of diameters in
ensembles of general configuration model networks, in both the supercritical
and the subcritical regimes, 
namely above and below the percolation transition, 
respectively
\cite{Bollobas2001,Newman2010}.
In the special case of subcritical ER networks it was found that the
distribution of diameters follows a Gumbel distribution
\cite{Hartmann2018}.

\section{Summary}

We derived a closed-form analytical expression for the DoE
of RRGs that consist of $N$ nodes of degree $c$.
The DoE is expressed in terms of the tail distribution
$P(E > \ell) \simeq 1 - \exp \left[ - \exp  \left( - \frac{ e^{b \ell} - \mu }{\beta} \right) \right]$,
where the distance $\ell$ takes integer values,
$b = \ln (c-1)$ is the shape parameter,
$\beta = \frac{c-2}{c} N$ 
is the scale parameter and
$\mu =   \frac{c-2}{c} N \ln N$
is the location parameter.
By providing the full distribution rather than a single characteristic length scale,
we present a detailed view of the large-scale structure.
In spite of the fact that the degrees of all the nodes are the same,
their eccentricities exhibit non-trivial variations.
Using the tail-sum formula 
and the Euler-Maclaurin expansion,
we obtained a closed-form expression for 
the mean eccentricity.
We also calculated the
mode of the DoE,
which exhibits a staircase profile as a function of the network size.
Interestingly, the mode is given by
$E_{\rm mode} ={\rm Round} \left(  \langle E \rangle  \right)$.
We also calculated the variance ${\rm Var}(E)$ and 
showed that it exhibits oscillations as a function of the
network size $N$.
These analytical results may serve as benchmarks for algorithmic approaches
to eccentricity calculations in large sparse networks
\cite{Li2019}. 
The eccentricities are important in practical applications 
such as broadcasting and global dissemination,
where the network performance is determined by the longest delay times.

{\appendix

\section{The probability that an RRG of size $N$ and degree $c$ will consist of two or more components}

In this Appendix we consider the probability that an RRG of size
$N$ and degree $c$ will consist of two or more components.
We denote by $T_c(N)$ the total number of labelled $c$-RRGs of size $N$,
and by $S_c(N)$ the number of labelled $c$-RRGs of size $N$ that consist
of a single connected component.
Clearly, for any choice of the network size $N$ and the degree $c$,
the inequality $S_c(N) \le T_c(N)$ is satisfied.
For a given degree $c$, the smallest possible RRG is of size $c+1$.
This network is a complete graph, which clearly consists of a single connected component.
This implies that $S_c(c+1) = T_c(c+1) =1$.

The probability $P_{\rm M}$ that a random instance of a $c$-RRG of size $N$ will
consist of multiple components is given by

\begin{equation}
P_{\rm M} = 1 - \frac{S_c(N)}{T_c(N)}.
\label{eq:PM0}
\end{equation}
 
\noindent
For sufficiently large values of $N$, the number of labelled $c$-RRGs can be
approximated by
\cite{Bender1978,Bollobas1980,Mckay1991,Evnin2024}

\begin{equation}
T_c(N) \simeq \frac{ (cN)! }{ (cN/2)! 2^{cN/2} (c!)^{N} } e^{ - \frac{c^2-1}{4} }.
\label{eq:TcN0}
\end{equation} 

\noindent
In case that an RRG of size $N$ consists of more than one component,
the most likely situation is that it splits into a large component
of size $N-c-1$ (which must be at least of size $c+1$) 
and a small component of size $c+1$.
Taking into account other ways to split an RRG of size $N$
would yield subleading corrections.
Therefore, the number $S_c(N)$ 
of labelled $c$-RRGs of size $N$ that consist of a single connected component
can be approximated by

\begin{equation}
S_c(N) \simeq T_c(N) -   \binom{N}{c+1} T_c(N-c-1) S_c(c+1).
\label{eq:ScN0}
\end{equation}

\noindent
Inserting $S_c(N)$ from Eq. (\ref{eq:ScN0}) into Eq. (\ref{eq:PM0})
we obtain

\begin{equation}
P_{\rm M}(N,c) \simeq \binom{N}{c+1} \frac{ T_c(N-c-1) S_c(c+1) }{ T_c(N) }.
\label{eq:PM1}
\end{equation}

\noindent
Inserting $T_c(N)$ from Eq. (\ref{eq:TcN0}) into Eq. (\ref{eq:PM1}) 
and using the Stirling approximation,
we obtain

\begin{equation}
P_{\rm M}(N,c) \simeq \frac{ (c!)^{c} }{ (c+1)   c^{ \frac{  c(c+1) }{2}  }} N^{ -  \frac{ (c+1)(c-2) }{2} }.
\label{eq:PM2}
\end{equation}

\noindent
Inserting $c=3$, $4$ and $5$ into Eq. (\ref{eq:PM2}),
we obtain 

\begin{equation}
P_{\rm M}(N,3) \simeq \frac{2}{27} N^{-2},
\end{equation}

\begin{equation}
P_{\rm M}(N,4) \simeq \frac{81}{1,280} N^{-5},  
\end{equation}

\noindent

\begin{equation}
P_{\rm M}(N,5) \simeq \frac{1,327,104}{9,765,625} N^{-9},
\label{eq:PM5}
\end{equation}

\noindent
respectively.

The simulation results presented in this paper are for RRGs consisting of nodes
of degree $c=5$. From Eq. (\ref{eq:PM5}) it is clear that for $c=5$ and for the network sizes
used in the paper, the probability that an RRG will consist of two or more
components is vanishingly small.
Indeed, in the construction of the network instances used in the simulations
we have not encountered even a single case of a network that consists of 
more than one component (and should be rejected).

}

\end{document}